%% file: prep.tex
\documentclass[]{article}
\usepackage{multicol}
\usepackage{apj}
\def\lesssim{\mathrel{\hbox{\rlap{\hbox{\lower4pt\hbox{$\sim$}}}\hbox{$<$}}}}
\def\gtrsim{\mathrel{\hbox{\rlap{\hbox{\lower4pt\hbox{$\sim$}}}\hbox{$>$}}}}
\newcommand {\h} {$h^{-1} \, Mpc \ $}
\newcommand {\hh} {$h^{-1} \, Mpc$}
\newcommand {\ks} {$km~s^{-1} \ $}
\newcommand {\kss} {$km~s^{-1}$}

\begin{document}

\vspace{15mm}
\begin{center}
\uppercase{
Evolution of the Internal Dynamics of  Galaxy Clusters}\\
\vspace*{1.5ex}
{\sc Marisa Girardi and Marino Mezzetti}\\
\vspace*{1.ex}
{\small
Dipartimento di Astronomia, Universit\`{a}
degli Studi di Trieste, Via Tiepolo 11, I-34131 Trieste, Italy\\
E-mail: girardi@ts.astro.it, mezzetti@ts.astro.it\\}
\end{center}
\vspace*{-6pt}

\begin{abstract}

We consider a sample of 51 distant galaxy clusters at $0.15\lesssim z
\lesssim 0.9$
($<z>\sim0.3$), each cluster having at least 10 galaxies with
available redshift in the literature.  We select member galaxies,
analyze the velocity dispersion profiles, and evaluate in a
homogeneous way cluster velocity dispersions and virial masses.

We apply the same procedures already recently applied on a sample of
nearby clusters ($z<0.15$, Girardi et al. 1998b) in order to properly
analyze the possible dynamical evolution of galaxy clusters. We remark
problems induced by the poor sampling and the small spatial extension
of the sampled cluster region in the computation of velocity
dispersion.

We do not find any significant difference between nearby and distant
clusters.  In particular, we consider the galaxy spatial
distribution, the shape of the velocity dispersion profile, and the
relations between velocity dispersion and X--ray luminosity and
temperature.  Our results imply little dynamical evolution in the
range of redshift spanned by our cluster sample, and suggest that the
typical redshift of cluster formation is higher than that of the
sample we analyze.


\vspace*{6pt}
\noindent
{\em Subject headings: }
galaxies: distances and redshifts - X-rays: galaxies - cosmology: 
observations.

\end{abstract}

\begin{multicols}{2}
\section{INTRODUCTION}

The knowledge of the properties of galaxy clusters and of their
possible evolution plays an important role in the study of large scale
structure formation constraining cosmological models (e.g., Henry et
al. 1992; Oukbir \& Blanchard 1992; Colafrancesco \& Vittorio 1994;
Eke, Cole, \& Frenk 1996).  The evolution of their statistical
properties have been  studied previously. In particular, there is no
evidence of evolution in the bulk of population of X--ray selected
clusters (out to $z\sim0.8$; e.g., Burke et al. 1997; Jones et
al. 1998; Rosati et al. 1998), with evidence for a negative evolution
of the X--ray luminosity function holding only for the brightest
objects (e.g., Gioia et al. 1990; Vikhlinin et al. 1998; Rosati et
al. 2000).  There is evidence of a mild evolution of the cluster X-ray
temperature function (out to $z\sim0.8$, Henry 1997; Donahue \& Voit
1999), and of somewhat larger evolution of the internal
velocity--dispersion function (Carlberg et al. 1997b; Borgani et
al. 1999). Other recent studies concern the relations between X--ray
properties or between X-ray and optical properties finding no evidence
of evolutions (out to $z\sim0.4-0.5$; e.g., Mushotzky \& Scharf 1997;
Borgani et al. 1999; Schindler 1999).  Moreover, no evidence of
evolution is found for other cluster properties, such as the iron
abundance (out to $z\sim0.8$ Mushotzky \& Loewenstein 1997) and the
core radius of the distribution of hot intracluster medium (Vikhlinin
et al. 1998).  All such signs of evidence suggest a low value for the
matter density parameter $\Omega_m$ (Carlberg et al. 1997b; Fan,
Bahcall, \& Cen 1997; Henry 1997; Borgani et al. 2000; see Mushotzky
2000 for a review).

However, the validity of these studies relies on our actual
understanding of the internal physics of both nearby and distant
clusters.  In particular, there is evidence that several clusters at
moderate/distant redshift ($z\sim0.2$ out to $z\sim1$)
are far from the state of
dynamical equilibrium suggesting that present observations are
reaching the epoch of cluster assembly.  For instance, it is claimed
that distant clusters often show discrepancy in determination of mass
estimates (e.g., Miralda-Escud\'e \& Babul 1995; Wu \& Fang 1996;
1997), where the problems concern, in particular, the cores of
clusters (e.g., Allen 1998; Wu, Fang, Xue 1998), and are probably due
to the lack of dynamical equilibrium or of spherical symmetry (e.g.,
Allen, Fabian, \& Kneib 1996; Girardi et al. 1997b). Moreover, direct
optical and X-ray observations show the strong elongation of some
distant clusters (e.g., Gioia et al. 1999).

In this framework, it is worth to analyze the internal dynamics of
distant clusters comparing, in particular, the results with those
obtained for nearby clusters.  Here we focus our attention on the
results as they come from the kinematical and spatial analysis of
cluster member galaxies.

As for nearby clusters (at redshift $z\lesssim 0.15$), available
results are based on very large samples, up to $\gtrsim 100$ clusters,
each with several galaxy redshifts available and treated in
homogeneous way: the ENACS (ESO Nearby Abell Cluster Survey, Katgert
et al. 1998) sample, and compilations collecting ENACS data and other
clusters from the literature (den Hartog \& Katgert 1996; Fadda et
al. 1996, and the following updating by Girardi et al. 1998b --
hereafter F96 and G98, respectively).  Significant substructures are
found for 30--$40\%$ of clusters from both the distribution of member
galaxies (e.g., Girardi et al. 1997a; Biviano et al. 1997; Solanes,
Salvador-Sol'e, and Gonz\'alez-Casado 1999) and X--ray analyses (Jones
\& Forman 1999), with a good one--to--one correspondence between the
optical and the X--ray images (Kolokotronis et al. 2000).  However,
with the exception of strongly 

\end{multicols}
\hspace{-13mm}
\begin{minipage}{20cm}
\renewcommand{\arraystretch}{1.2}
\renewcommand{\tabcolsep}{1.2mm}
\begin{center}
TABLE 1\\
{\sc Cluster Sample \\}
\footnotesize 
\input{tab1}
\end{center}
\vspace{-2mm}
\input{comm_tab1}
\end{minipage}
\begin{multicols}{2}

\noindent substructured clusters (e.g., a $10\%$
of bimodal clusters, see Girardi et al. 1997a; G98), most clusters
seem not to be far from a global dynamical equilibrium. Finally,
galaxy light is a good tracer of dark matter (e.g., Natarajan et
al. 1998).  The comparison between reliable estimates of velocity
dispersions and X--ray temperatures of clusters suggests that the
galaxy and hot gas components are not far from energy equipartition
per unit mass; the possible discrepancies are likely to require
extra--heating sources for poor clusters (e.g., White 1991; Bird,
Mushotzky, \& Metzler 1995; G98).  There is an overall agreement
between mass estimates inferred from the analysis of member galaxies
and those from X--ray analysis of the hot gas (G98). Other detailed
studies concern comparative analyses of different galaxy populations
and their use as tracers of the cluster potential (e.g., early-- and
late--type ones, galaxies with or without emission lines, see, e.g.,
Biviano et al. 1997; Adami, Biviano, \& Mazure 1998).

As for more distant clusters ($z>0.2$), the possible evolution of
member galaxies is well studied (e.g., Butcher \& Oemler 1978;
Dressler et al. 1997; Abraham et al. 1998), but there are less
definitive results on cluster internal dynamics. Most results come from
the analysis of the 16 clusters at intermediate redshifts of CNOC
(Canadian Network for Observational Cosmology, $0.18<z<0.55$ ; Yee,
Ellingson, \& Carlberg 1996b) which represent a remarkably
homogeneous sample.  In particular, as also found in nearby clusters,
Lewis et al.  (1999) claim for consistency between masses coming from
optical and X--ray data, and Carlberg et al. (1997a) find that blue
and red galaxies have different distributions in velocity and
position.  However, the difficulty of obtaining many redshifts in
distant clusters has prevented from building larger samples. Rather,
several works, concerning one or a small number of clusters, and using
different techniques of analysis, can be found in the literature.

The availability of a variety of techniques, already applied to nearby
clusters, suggests their application to distant clusters.  We thus
ensure the homogeneity of our results over a large range of
cosmological distances.  A homogeneous analysis is in fact fundamental
for the understanding of the evolution of cluster properties.

Here, we apply the techniques already used by G98 (cf. also F96) on a
sample of 170 nearby clusters (at $z<0.15$, data from ENACS and other
literature) to analyze a collection of 51 distant clusters at
$0.15\lesssim z\lesssim 0.9$.

The paper is organized in the following manner.  We shortly describe
the data sample and our selection procedure for cluster membership
assignment in \S~2 and \S~3, respectively.  We compute internal
velocity dispersions and masses for clusters in \S~4, with the
exception of the three clusters with strong dynamical uncertainties
which are discussed in the Appendix~A. We compare the ``active'' and the
non ``active'' galaxies in \S~5.  We compare our results with those
coming from X--ray and weak gravitational lensing analyses in \S~6.
We give a brief summary of our main results and draw our conclusions
in \S~7.

Unless otherwise stated, we give errors at the 68\% confidence
level (hereafter c.l.)

A Hubble constant of 100 $h$ \ks $Mpc^{-1}$ and a deceleration
parameter of $q_0=0.5$ are used throughout.

\section{THE DATA SAMPLE}

We consider 51 clusters at moderate/distant redshift $z>0.15$
(median
$z=0.33$), each cluster having at least 10 galaxies with available
redshift and showing a significant peak in the redshift space.  We 
remark
that, due to the the difficulty of obtaining redshift data
for distant clusters, we relax the requirements already applied to the
sample of nearby clusters of G98 ($z<0.15$, cf. also F96).  In
particular, we consider also clusters with less than 30 available
galaxy redshift, although we never take into account clusters which,
after the procedure of the rejection of interlopers, are left with $<
5$ member galaxies. We relax also other requirements concerning the
reliability of the estimate of 
velocity dispersion, i.e.  the
requirement of small errors on velocity dispersion ($\lesssim 150$
\kss), and of flat integrated velocity dispersion profile in external
cluster regions (cf. \S~4.1 for more details).

Cluster data are collected from the literature.  In order to achieve
sufficiently homogeneous cluster data, the galaxy redshifts in each
cluster are usually taken from only one reference source; different
sources are used only when the data-sets are proved to be compatible.
The data used for each cluster concern galaxy positions, redshifts
with the respective errors, and, when available,
spectral/morphological information.

Table~1 lists all the 51 clusters considered: in Col.~(1) we list the
cluster names; in Col.~(2) we report other alternative names found in
the literature; in Col.~(3) the number of galaxies with measured
redshift in each cluster field; in Col.~(4) the data references.

\section{CLUSTER MEMBER SELECTION}
In order to select member galaxies, we apply the same procedure as G98
(cf. also F96).

We first use position and velocity information sequentially; then we
use the two sets of data combined.  In the first two steps we  use
the adaptive kernel technique by Pisani (1993,1996) as described in the
Appendix A of Girardi et al. (1996).  The adaptive kernel technique is a
nonparametric method for the evaluation of the density probability
function underlying an observational discrete data set.  For each
detected peak, the method gives the corresponding significance and
object density, as well as the associate objects.

Firstly, we apply to each cluster field the two-dimensional adaptive kernel
analysis to detect clusters which show an obvious bimodality in their
projected galaxy distribution: i.e. formed by two significant ($>99\%$
c.l.) clumps separated by a distance of $\gtrsim 0.5$ \h. These clumps 
are then analyzed separately.  

Afterwards, we apply the one-dimensional analysis to find the significant
peaks in velocity distributions.  The 

\end{multicols}
\vspace{-20mm}
\begin{minipage}{20cm}
\renewcommand{\arraystretch}{1.2}
\renewcommand{\tabcolsep}{1.2mm}
\begin{center}
\vspace{-3mm}
TABLE 2\\
{\sc Cluster Membership \\}
\footnotesize 
\input{tab2}
\end{center}
\vspace{-3mm}
\hspace{45mm}
\input{comm_tab2}
\end{minipage}
\begin{multicols}{2}

\noindent main cluster body is 
naturally identified as the highest significant peak. All galaxies not
belonging to this peak are rejected as non cluster members.  F96 and
G98 required that peaks are significant at the $99\%$ c.l.  and, for
clusters with secondary peaks, they assumed that the peaks are
separable when their overlapping is $\le 20 \%$ and their velocity
separation is $\Delta v\ge$ 1000 \ks (here, we consider $1000$ \ks in
the appropriate cluster rest--frame).  In dealing with distant
clusters, we apply the peak analysis to very poor samples, so
obtaining small peak probability: in a few fields we identify
clusters with the highest peak having significance $<99\%$ (but always
$>95\%$).

The combination of position and velocity information, represented by
plots of velocity vs. clustercentric distance, reveals the presence
of surviving interlopers (e.g., Kent \& Gunn 1982; Reg\"os \& Geller
1989).  To identify these interlopers in the above-detected systems we
apply the procedure of the ``shifting gapper'' by F96.  We apply the
fixed gap method to a bin shifting along the distance from the cluster
center. According to F96 prescriptions, we use a gap of $\ge 1000$ \ks
(in the cluster rest--frame) and a bin of 0.4 \hh, or large enough to
include 15 galaxies. As for very poor distant clusters (with less than
15 members), we reject galaxies that are too far in velocity from the
main body of galaxies of the whole cluster (considering a somewhat larger
gap,  cf. \S~3.1 for details).

When early- and late-galaxy type populations showed different mean
and variance in the velocity distribution, Girardi et al. (1996)
retained only the early population as better tracer of the cluster
potential. Similarly, when only spectral information was available,
F96 applied the same procedure by rejecting emission line galaxies
(hereafter ELGs, while NELGs indicate galaxies without emission
lines). Indeed there is evidence that ELGs lead to too high estimates
of internal velocity dispersion (e.g., Koranyi \& Geller 2000) and
that they could be not in dynamical equlibrium within the cluster
potential (Biviano et al. 1997).  For distant clusters, galaxy
morphologies are generally not available, but spectral type and/or
color give information about the presence of some nuclear galaxy
activity or strong star formation.  For the 43 out of 51 cluster
samples with available information, we reject the likely ``active''
galaxies (hereafter AGs), i.e. galaxies having a strong star formation
activity and/or signs of nuclear activity (see our classification in
\S~3.2).

Table~2 lists the results of the member selection procedure
(cf. \S~3.1 for details on some specific clusters). In Col.~(1) we
list the system name, i.e.  the name of the parent cluster with
possible indication of the peak (e.g. A1689a and A1689b); in Col.~(2)
the number of galaxies found by the adaptive kernel method in each
peak, $N_p$; in Col.~(3) the number of galaxies left after the
``shifting gapper'', $N_g$; in Col.~(4) the number of member galaxies
after the rejection of the AGs, $N_m$, and used to compute the mean
redshift determined via the biweight estimator (Beers, Flynn, \&
Gebhardt 1990), and the cluster center as determined via the
two--dimensional adaptive kernel (in Cols.~(5) and (6), respectively).

\subsection{Results for Specific Clusters}

According to our analysis of galaxy distribution, we find indication
for bimodality only in the case of A115 (cf. also Forman et al. 1981;
Beers, Geller, \& Huchra 1983). Moreover, we consider only the Southern
peak, A115S, since the Northern peak has too few galaxies to survive to
the whole procedure of member selection.

According to our analysis of cluster velocity distributions, out of
the 51 cluster fields here analyzed, we find 51 well--separated peaks
(45 from 45 one-peaked fields and six from three two-peaked fields) and three
fields with two strongly superimposed peaks.

The three fields showing two separable peaks are A3889, CL0949+44
(cf. also Dressler \& Gunn 1992), and CLJ0023+0423.  In particular,
the field of the CLJ0023+0423 cluster shows a complex structure in the
velocity distribution, containing a system of four peaks strongly
superimposed; however, when reanalyzing only galaxies belonging to
this system, we find two peaks.  These peaks corresponds to those
found by Lubin, Postman, \& Oke (1998a, CLJ0023+0423 ``A'' and ``B''
instead of our ``b'' and ``a'').  

The cluster fields for which the peak separation is not secured at a
high c.l.  are: A1689 (cf. also Girardi et al. 1997), A2744, and
A3854. The strongly overlapped peaks could indicate the presence of
substructures in a single system and, in this case, the dynamics of
these clusters is strongly uncertain; therefore we consider both the
case with the two peaks disjoined or together (e.g., A1689a, A1689b,
and A1689ab), cf. the Appendix A for other details.

As for very poor distant clusters (with less than 15 galaxies), the
procedure of the ``shifting gapper'', which works with a gap of $1000$
\ks in a shifting bin considering at least 15 galaxies, cannot be
applied. In these cases we reject galaxies that are too far in
velocity from the main body of galaxies of the whole cluster rather
than in a shifting bin.  Moreover, we adopt a slightly larger gap
since the suitable size of the gap increases with the available
statistics (cf. the density-gap by Adami et al.  1998).  We reject one
galaxy in CL0054-27, one galaxy in J2175.23C, and two galaxies in
1E0657-56, where the gap is $\gtrsim 2000$ \kss.  The situation is
less obvious for other two galaxies in CL0054-27 and other three
galaxies in J2175.23C, where the respective gaps are $\sim 1150$ \kss,
$\sim 1300$ \kss, respectively.  For CL0054-27, the two uncertain
members are close to the center and we decide to retain
them. Instead, for J2175.23C, we decide to reject the three uncertain
members,  which are connected to the main body of
galaxies only thanks to the presence of an AG. Moreover, two of the
uncertain members are AGs, and the other uncertain member would result
the most distant galaxy from the cluster center.  Monte Carlo
simulations performed in \S~4.4 show that, in the case of poor
statistics, a fixed gap of $\sim1250$ \ks allows us to well recover,
on average, the estimate of velocity dispersion.

As for the combined analysis of position and velocity information, the
plot of rest--frame velocity versus projected clustercentric distance
of MS1358.4+6245 shows the existence of a close system corresponding
to a southern group (cf.  Carlberg et al. 1996).  We exclude this
group rejecting all galaxies outside $1.2$ \h (cf. also Borgani et
al. 1999).

\subsection{Classification of ``Active Galaxies''}

As for our classification of AGs, in several cluster samples only main
galaxy spectral features are reported. In these cases we classify as
``active'' galaxies to be rejected those where the presence of
emission lines is reported.  For other clusters, where more detailed
information is given, we always reject galaxies with very strong
emission lines or classified as starbust or AGN: in the following we
describe the classification adopted for these specific studies.

As for the data by Postman, Lubin, \& Oke (1998), we reject galaxies with the
presence of a $[OII]$ line with an equivalent width of $EW[OII]\gtrsim
15$ \AA $\;\;$, which corresponds, according to the authors, to an
active, star forming galaxy.  

As for the data by Dressler et al. (1999), we consider as AGs those galaxies
classified from their spectra as ``e(a)'' (with strong Balmer
absorption plus [OII] emission), ``e(n)'' (having AGN spectra), or
``e(b)'' (with very strong [OII] emission, possibly starbust
galaxies).  As for data by Dressler \& Gunn (1992), ``active'' galaxies
are those classified with ``e'' (with emission lines, usually [OII] or
[OIII]) or ``n'' (with very strong emission, likely due to AGN).

As for the CNOC clusters (Yee et al. 1996a; Ellingson et
al. 1997; Abraham et al. 1998; Ellingson et al. 1998; Yee et
al. 1998), we consider as AGs those galaxies classified from their spectra
with ``5'' (with emission lines, likely irregular galaxies) or ``6''
(likely AGN/QSO). 

Finally, we also consider as AGs  those galaxies labeled ``starbust'' by Couch
et al. (1998, and previous works by Couch \& Sharples 1987 and Couch
et al. 1994), which are characterized by blue colors and
emission-filled H$\delta$ lines.

\section{ANALYSIS OF \\
CLUSTER INTERNAL DYNAMICS}

The analysis of the three clusters which show strongly superimposed
peaks (A1689, A2744, and A3854) is postponed to the Appendix~A.  In the
following sections we analyze the 45 cluster fields which show only
one peak in their velocity distribution and the three cluster fields
which show two separable peaks for a total of 51 well--defined systems
(cf. \S~3.1).  This sample
can be compared to that of 160 well--separated peaks for nearby clusters of
G98. We remark that, on average, the distant clusters are less well
sampled as for both  the number of cluster members (median $N_m=21$
vs. 39), and  the spatial extension $R_{max}$, which is the
clustercentric distance of the most distant galaxy from the cluster
center (median $R_{max}=0.64$ vs. 1.45 \hh).  Throughout this analysis
we apply homogeneous procedures already used by G98 (cf. also F96).

\subsection{Velocity Dispersions}
We estimate the ``robust'' velocity dispersion line--of--sight,
$\sigma_v$, by using the biweight and the gapper estimators when the
galaxy number is larger and smaller than 15, respectively (cf. ROSTAT
routines -- see Beers et al.  1990), and applying the relativistic
correction and the usual correction for velocity errors (Danese, De
Zotti, \& di Tullio 1980).  In particular, for a few cases where the
velocity error is not available, we assume a typical velocity error of
$300$ \kss.  When the correction for velocity errors leads to a
negative value of $\sigma_v$, we list $\sigma_v=0$.

Following F96 (cf. also Girardi et al. 1996) we analyze the
``integral'' velocity dispersion profile (hereafter VDP), where the
dispersion at a given (projected) radius is evaluated by using all the
galaxies within that radius, i.e. $\sigma_v(<R)$. The VDPs allow to
check the robustness of $\sigma_v$ estimate. In particular, although
the presence of velocity anisotropy in galaxy
orbits  can strongly influence the value
of $\sigma_v$ computed for the central cluster region, it does not
affect the value of the $\sigma_v$ computed for the whole cluster
(e.g., Merritt 1988). The VDPs of nearby clusters show strongly
increasing or decreasing behaviors in the central cluster regions, but
they are flattening out in the external regions (beyond $\sim $1 \hh,
cf. also den Hartog \& Katgert 1996) suggesting that in such regions
they are no longer affected by velocity anisotropies.  Thus, while the
$\sigma_v$-values computed for the central cluster region could be a
very poor estimate of the depth of cluster potential wells, one can
reasonably adopt the $\sigma_v$ value computed by taking all the
galaxies within the radius at 
which the VDP becomes roughly constant.
As for the distant clusters we analyze, when the data are good enough,
the VDPs show a behavior similar to that of nearby clusters
(cf. Figure~1).  Unfortunately, distant clusters suffer for the poor
sampling, and also for the small spatial extension of the sampled
cluster region.  Indeed, the strongly decreasing VDP in the external
sampled regions of some clusters (maybe the striking cases are AS506,
CL0017-20, CL0054-27, 3C295) suggests that the correct estimates of
velocity dispersions could be smaller than those, $\sigma_v$, we can
estimate with present data; therefore, in these cases, $\sigma_v$
should be better interpreted as an upper limit (see also some cases in
F96; these cases were then not considered in G98). In other cases,
when the member galaxies are too few, the analysis of VDPs does not
allow any conclusion.

In Table~3 we report the value of $\sigma_v$ computed considering all
member galaxies. However, we make a note for the clusters which do not
share the requirements for the nearby clusters of F96 and G98, i.e.:
with an original number of galaxies in the field smaller than 30; with
a peak in the velocity distribution less significant than $99\%$; with
an error on $\sigma_v$ larger than 150 \kss; with a VDP which is
poorly defined or without a flat behavior in the external cluster
regions.

After fixing the cosmological background, the theory of a spherical model
for nonlinear collapse allows to recover the value of the radius of
virialization, $R_{vir}$, within which the cluster can be considered not far
from a status of dynamical equilibrium.  The relation between the density
of a collapsed (virialized) region and the cosmological density is
$\rho_{vir}= 18\pi^2 \rho_{cr}= 18 \pi^2 \times 3 H^2/8\pi G$ (for a
$\Omega_m=1$ universe).  As a first approximation, the mass contained
within $R_{vir}$, $M_{vir}=(4\pi/3)\cdot R_{vir}^3\rho_{vir}$, is given
by the virial estimate $(3\pi /2) \cdot (\sigma_v^2 R_{vir}
f_{\Sigma}/G)$, where $f_{\Sigma}$ depends on the details of galaxy
spatial distribution (e.g., G98).  Therefore
$R_{vir}^2=\sigma_v^2 f_{\Sigma}/(6\pi H^2)$, where $H=100\cdot h
(1+z)^{3/2}$.  For nearby clusters G98 give a first roughly estimate of

\includegraphics{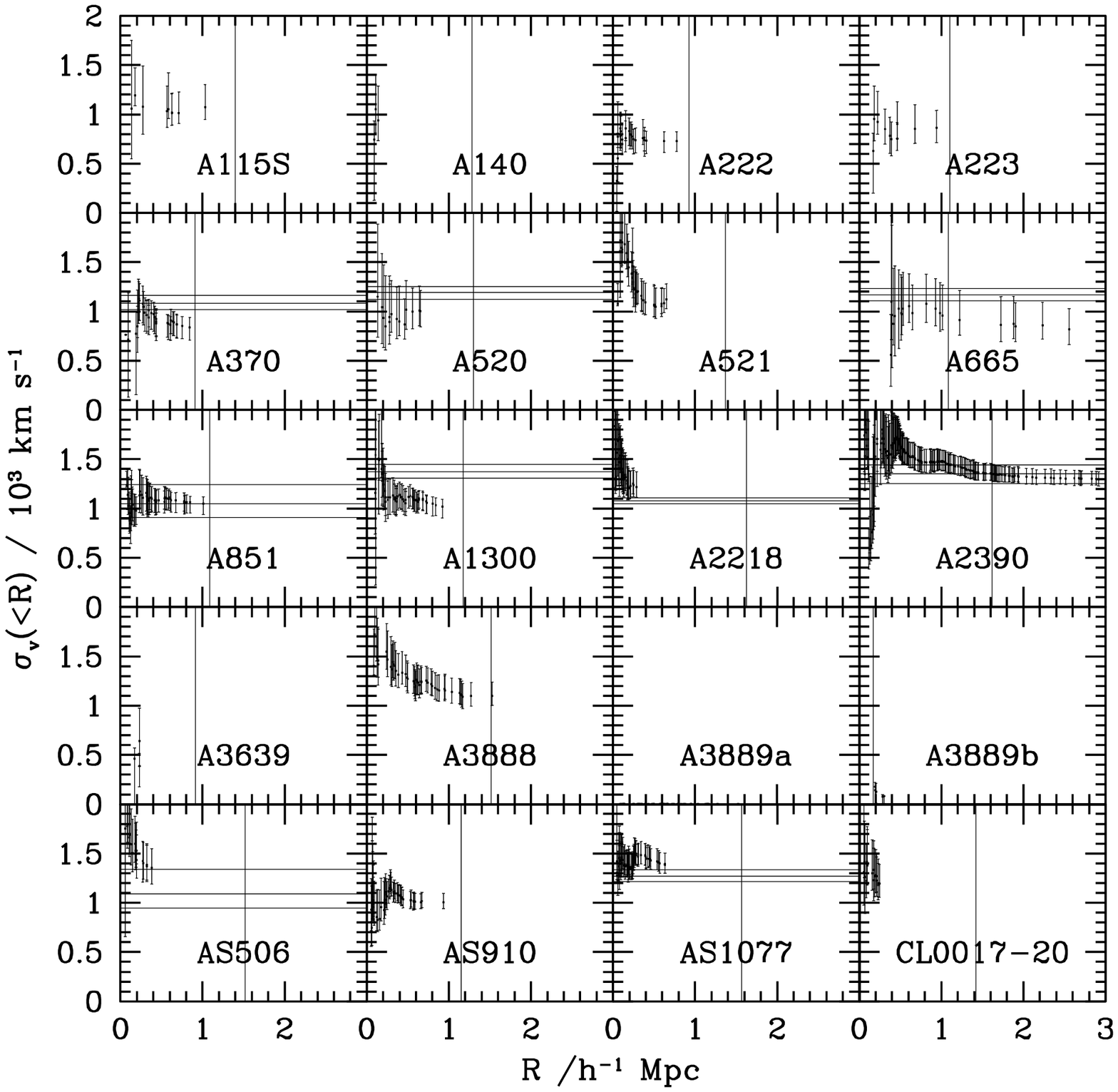}
$\ \ \ \ \ \ $\\
\vspace{8.2truecm}
$\ \ \ $\\
\vspace{0.5cm}

\includegraphics{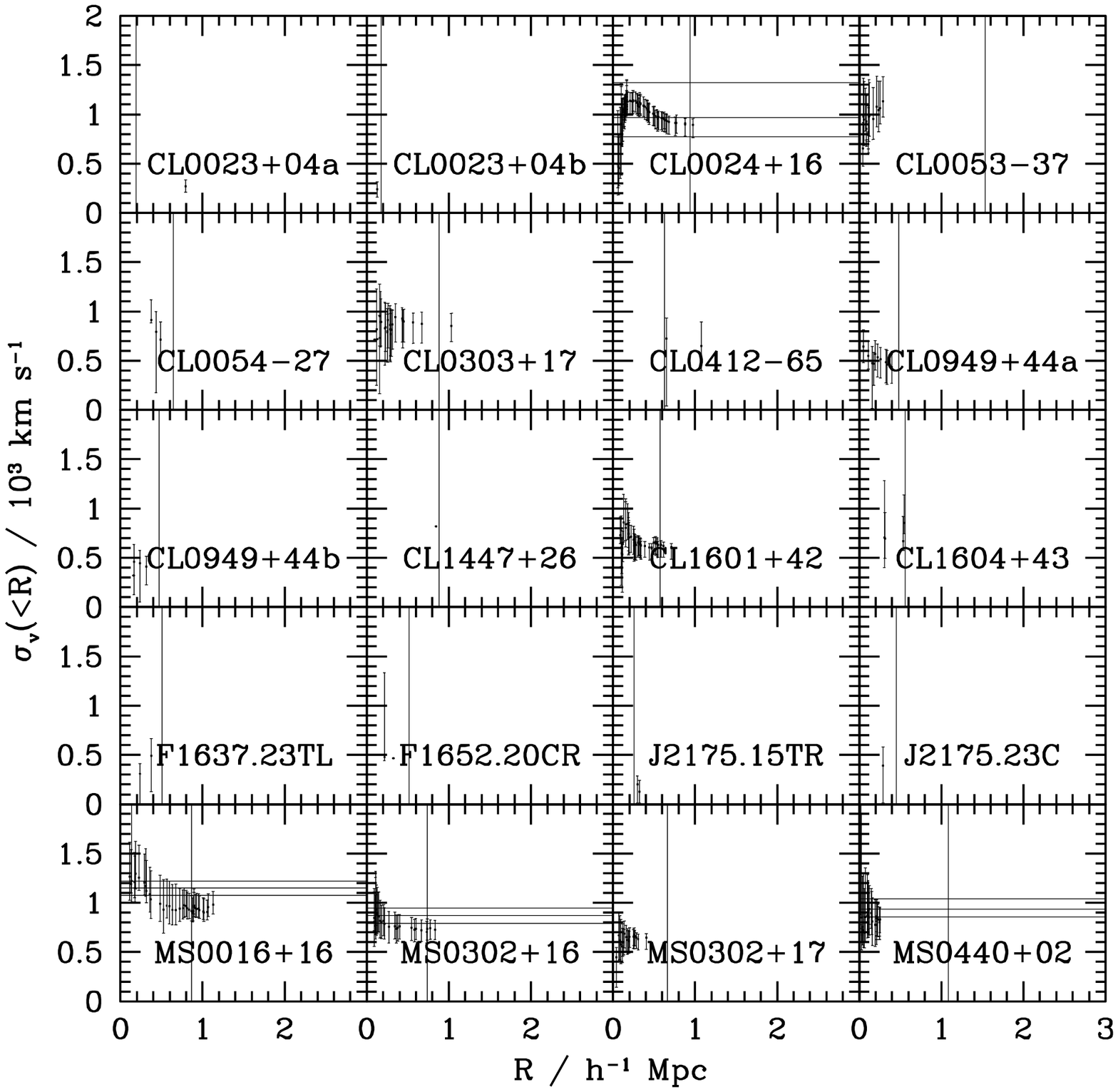}
$\ \ \ \ \ \ $\\
\vspace{8.2truecm}
$\ \ \ $\\


\vspace{0.5cm}

\includegraphics{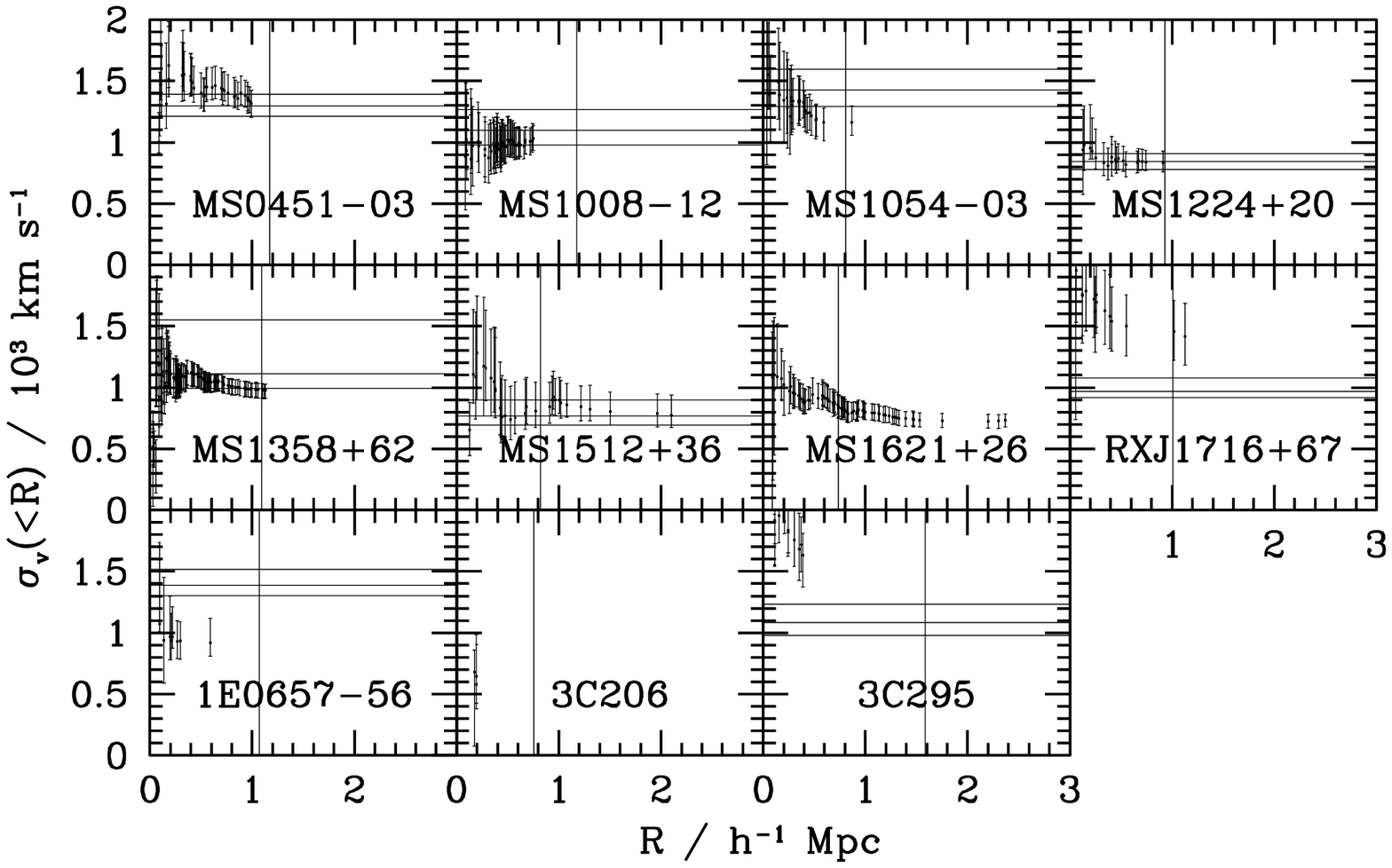}
$\ \ \ \ \ \ $\\
\vspace{4.truecm}
$\ \ \ $\\
{\small\parindent=3.5mm {Fig.}~1.---
Integrated line--of--sight velocity dispersion
profiles $\sigma_v(<R)$, where the dispersion at a given (projected)
radius from the cluster center is estimated by considering all
galaxies within that radius. The bootstrap error bands at the $68\%$
c.l. are shown.  The horizontal lines represent X--ray temperature
with the respective errors (cf. Table~4) transformed in $\sigma_v$
imposing $\beta_{spec}=1$ (where $\beta_{spec}=\sigma_v^2/(kT/\mu
m_p)$, with $\mu$ the mean molecular weight and $m_p$ the
proton mass).  The vertical faint lines indicate the virialized region
within $R_{vir}$.
}
\vspace{5mm}

\noindent $R_{vir}\sim 0.002 \cdot \sigma_v$ ($km^{-1}s\ $ \hh). A following
re--estimate of Girardi et al. (1998a) suggests rather a scaling factor of
0.0017. Since we find that distant clusters have a galaxy distribution
similar to that of nearby ones (see in the following), we adopt here the
same scaling relation with $\sigma_v$: 
\begin{equation}
R_{vir}\sim 0.0017 \cdot \sigma_v/(1+z)^{3/2} \,\, (km^{-1} s\  h^{-1}\,Mpc) 
\end{equation}
\noindent introducing only the scaling with redshift (cf. also Carlberg et
al. 1997c for a similar relation).

\subsection{Galaxy Distribution}
As for the study of the spatial distribution of galaxies
within distant clusters,
following G98 (cf. also Girardi et al. 1995; Adami et al. 1998) we fit
the galaxy surface density of each cluster to a King--like
distribution (comparable to the $\beta$-profile 
in fitting the distribution of hot diffuse intracluster 
gas):
\begin{equation}
\Sigma(R)=\frac{\Sigma_0}{[1+(R/R_c)^2]^{\alpha}} \ ,
\end{equation}
\noindent where $R_c$ is the core radius and $\alpha$ is the parameter
which describes the galaxy distribution in external regions
($\alpha=1$ corresponds to the classical King distribution).  This
surface density profile corresponds to a galaxy volume-density
$\rho=\rho_0/[1+(r/R_C)^2]^{3\beta_{fit,gal}/2}$, with
$\beta_{fit,gal}=(2\alpha+1)/3$, i.e.  $\rho(r) \propto r^{-3
\beta_{fit,gal}}$ for $r>> R_C$.  We perform the fit through the
Maximum Likelihood technique, allowing $R_C$ and $\alpha$ to vary from
0.01 to 1 \h and from 0.5 to 1.5, respectively. 

To avoid possible effects due to the non circular sampling of
clusters, by visual inspection of the original sampled region of each
cluster we extract the largest circular region, with center as in
Table~2, there inscribed. We perform the fit within this circular
cluster region whose radius we define as $R_{max,c}$.  We consider
only the 30 clusters with at least ten member galaxies within
$R_{max,c}$ and, in particular, a subsample of 13 clusters with
$R_{max,c}/R_{vir}>0.5$.  

The median value of $\alpha$, with the respective errors at the $90\%$
c.l., is $=0.63_{-0.08}^{+0.08}$, and 0.67 is found when we consider
only the 13 clusters with a large sampled radius.  This value agrees
with $\alpha=0.70_{-0.03}^{+0.08}$ found for nearby clusters, and
corresponds to a $\beta_{fit,gal}\sim0.8$, i.e. to a volume
galaxy--density $\rho \propto r^{-2.4}$.  After fixing $\alpha=0.7$,
we again fit the galaxy distribution of each cluster, obtaining a
median value of $R_c=0.045_{-0.015}^{+0.005}$ \h (and 0.05 \h for the
well--sampled 13 clusters). Thus, in our cluster sample, the typical
value of $R_c$ (and $R_{vir}/R_c \sim 20$) is again in agreement with
that found in nearby clusters where $R_c=0.05\pm0.01$ \hh.  Hereafter,
we assume the above King--modified distribution, with the same
parameters of nearby clusters, i.e.  $\alpha=0.7$ and $R_c=0.05$ \hh,
for all clusters of our sample.

\subsection{Virial Masses and Velocity Anisotropies}
Assuming that clusters are spherical, non rotating systems, and that
the internal mass distribution follows galaxy distribution, cluster
masses can be computed throughout the virial theorem (e.g., Limber \&
Mathiews 1960; The \& White 1986) as:
\begin{equation}
M=M_V-C=\frac{3\pi}{2} \cdot \frac{\sigma_v^2 R_{PV}}{G}-C,
\end{equation}
\noindent where the projected virial radius,
\begin{equation}
R_{PV}=N(N-1)/(\Sigma_{i> j} R_{ij}^{-1}),
\end{equation}
\noindent describes the galaxy
distribution and is computed from projected mutual galaxy distances,
$R_{ij}$; $C$ is the surface term correction to the standard virial
mass $M_V$ and it is due to the fact that the system is not entirely
enclosed in the observational sample (cf. also Carlberg et al. 1996;
G98).

Following G98 we want to estimate cluster masses contained within the
radius of virialization, $R_{vir}$.  In fact, clusters cannot be
assumed in dynamical equilibrium outside $R_{vir}$ and considering
small cluster regions leads to unreliable measures of the potential
($\sigma_v$ could be strongly affected by velocity anisotropies) and
of the surface term correction (Koranyi \& Geller 2000).  

Only few distant clusters are sampled out to $R_{vir}$.  
As for $\sigma_v$, the above analysis of the VDP give indications
about its reliability, i.e. VDPs which are flat in the external
cluster regions will give reliable estimates of $\sigma_v$ even when
clusters are not sampled out to $R_{vir}$.  As for $R_{PV}$, which
describes the galaxy spatial distribution, it can be recovered in an
alternative theoretical way from the knowledge of the parameters of
the King--like distribution (Girardi et al. 1996; see also G98 for a
simple analytical approximation in the case of $\alpha=0.7$ and
$R_c/R_{vir}=0.05$).  This procedure allows to compute $R_{PV}$ at
each cluster radius and, in particular, we compute $R_{PV}$ at
$R_{vir}$, which is needed in the computation of the mass within
$R_{vir}$.  By using well--sampled nearby clusters G98 verified the
reliability of this alternative estimate and evaluated the typical
error introduced by the use of the average King--like parameters of
the sample ($\sim 0.2$ \h, corresponding to about $25\%$ of $R_{PV}$).

The computation of the $C$ correction at the boundary radius, $b$,
\begin{equation}
C=M_V \cdot 4 \pi b^3 \frac{\rho(b)}{\int^{b}_0 4\pi r^2\rho dr}\left[\sigma_r(b)/\sigma(<b)\right]^2,
\end{equation}
\noindent requires the knowledge of the velocity anisotropy of galaxy
orbits.  In fact, $\sigma_r(b)$ is the radial component of the
velocity dispersion $\sigma(b)$, while $\sigma(<b)$ refers to the
integrated velocity dispersion within $b$; here $b=R_{vir}$. Having
assumed that in clusters the mass distribution follows the galaxy
distribution, one can use the Jeans equation to estimate velocity
anisotropies from the data, i.e. from the (differential) profile of
the line--of--sight velocity dispersion, $\sigma_v(R)$. Unfortunately,
this profile requires a large number of galaxies and we can compute it
only by combining together the data of many clusters, without
preserving cluster individuality.

\includegraphics{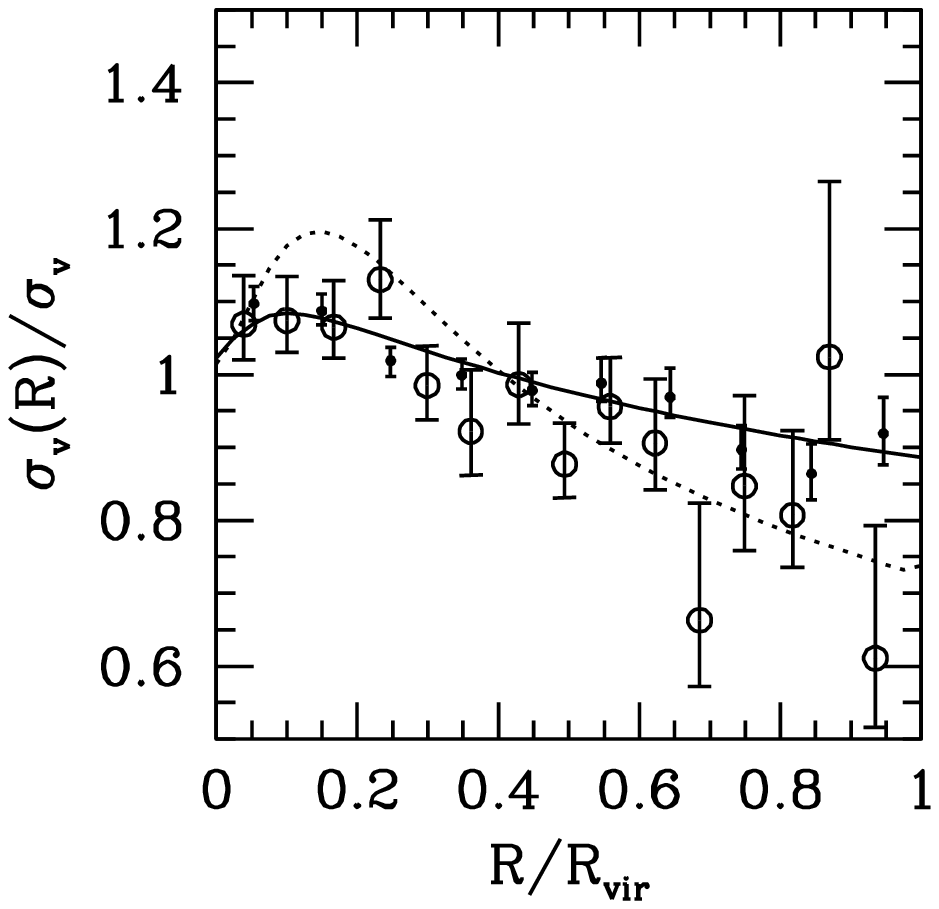}
$\ \ \ \ \ \ $\\
\vspace{7truecm}
$\ \ \ $\\
{\small\parindent=3.5mm {Fig.}~2.---
The (normalized) line--of--sight velocity
dispersion, $\sigma_v(R)$, as a function of the (normalized) projected
distance from the cluster center.  The points represent data combined
from all clusters and binned in equispatial intervals. We give the
robust estimates of velocity dispersion and the respective bootstrap
errors.  We give the results for distant clusters (open circles) and
for nearby clusters taken from Girardi et al. (1998b, filled circles).
The solid and dotted line represent the models for isotropic and
moderate radial orbits of galaxies, respectively (see text).
}
\vspace{5mm}

For both nearby and distant clusters, Figure~2 shows the observational
$\sigma_v(R)$ computed by combining together the galaxies of all
clusters, i.e.  by normalizing distances to $R_{vir}$ and velocities,
relative to the mean cluster velocity, to the observed global velocity
dispersion $\sigma_v$. For nearby 

\end{multicols}
\hspace{-9mm}
\begin{minipage}{20cm}
\renewcommand{\arraystretch}{1.2}
\renewcommand{\tabcolsep}{1.2mm}
\begin{center}
TABLE 3\\
{\sc Dynamical Properties \\}
\footnotesize 
\input{tab3}
\end{center}
\input{comm_tab3}
\end{minipage}
\begin{multicols}{2}

\noindent clusters the observational profile
is well described by a theoretical profile obtained by the Jeans
equation, assuming that velocities are isotropic, i.e. that the
tangential and radial components of velocity dispersion are equal
(i.e., the velocity anisotropy parameter ${\cal
A}=1-\sigma_{\theta}^2(r)/\sigma_r^2(r)=0$).  
For distant clusters
this model is less satisfactory, but cannot be rejected
being acceptable at the $\sim15\%$ c.l. (according to the $\chi^2$
probability).

In order to give $C$--corrections more appropriate to each individual
cluster G98 used a profile indicator, $I_p$, which is the ratio
between $\sigma_v(<0.2\times R_{vir})$, the line--of--sight velocity
dispersion computed by considering the galaxies within the central
cluster region of radius $R=0.2\times R_{vir}$, and the global
$\sigma_v$. According to the values of this parameter, they divided
clusters in three classes containing the same number of clusters:
``A'' clusters with a decreasing profile ($I_p>1.16$), ``C'' clusters
with a increasing profile ($I_p<0.97$), and an intermediate class
``B'' of clusters with very flat profiles ($0.97<I_p<1.16$).  Each of
the three types of profiles can be explained by models with a
different kind of anisotropy, i.e. radial, isotropic, and circular
orbits in the case of A, B, and C clusters, respectively, requiring
different values of $[\sigma_r(R_{vir})/\sigma(<R_{vir})]^2$, cf. G98.
We can define 14, 11, and 8 clusters of class A, B, and C,
respectively, and for each class we use the respective
$[\sigma_r(R_{vir})/\sigma(<R_{vir})]^2$ given by G98 to determine the
$C$--corrections.  The median values of the relative corrections
$C/M_V$ are $45\%$, $20\%$, and $14\%$ for A, B, and C clusters,
respectively.  For 18 clusters we cannot define the kind of profile
and we assume the intermediate one.  The median correction on the
whole sample is then $C/M_V\sim21\%$, very similar to that found by
G98 for nearby clusters and to that suggested by Carlberg et
al. (1997c) for CNOC clusters.
 
In Table~3 we list the results of the cluster dynamical analysis: the
number of cluster members as taken from Table~2, $N_m$ (Col.~2); the
clustercentric distance of the most distant galaxy from the cluster
center, $R_{max}$ (Col.~3); the global line--of--sight velocity
dispersion $\sigma_v$ with the respective bootstrap errors (Col.~4);
the radius which defines the region of virialization, $R_{vir}$
(Col.~5); the projected virial radius, $R_{PV}$, computed at $R_{vir}$
(Col.~6); the cluster type according to their velocity dispersion
profile, $T$ (Col.~7); the estimate of cluster mass contained within
$R_{vir}$ as determined from the standard virial theorem, $M_V$, and
after the pressure surface term correction, $M$ (Cols.~8 and 9,
respectively). The errors on $M_V$ take into account of the errors on
$\sigma_v$ and the above quoted error of $25\%$ on $R_{PV}$.  The
percent errors on $M_{CV}$ are the same as for $M_V$, i.e. we neglect
uncertainties on $C$--correction.

There are some possible suggestions that galaxy orbits in distant
clusters have a somewhat more radial velocity anisotropy than those in
nearby clusters.  Figure~2 shows that the velocity dispersion profile
for distant clusters seems less flat than that of nearby clusters.
Indeed, according to the available data for distant clusters, also
models with moderate radial orbits, i.e.  ${\cal A}=r^2/(r^2+r_a^2)$
with $r_a=0.25 \times R_{vir}$, can be acceptable (at the $\sim 4\%$
c.l.).  Moreover, the distant clusters with strongly decreasing
profiles (type A) are many more than clusters with increasing profiles
(type C), while, for definition, the number of A and C types are equal
among nearby clusters. However, we verify that there is no significant
difference between the combined profiles of distant and nearby
clusters, and that distant and nearby clusters are not different
according to the median value of the profile parameter $I_p$.
Therefore, with present data , we conclude that the possible evidence
of a larger amount of radial velocity anisotropies in galaxy orbits of
more distant clusters is not significant.

\subsection{Robustness of the Results}

Here we address  the effects of the poor sampling
on our results, in particular on our estimates of velocity dispersions.

In order to check the effects of the small number of redshifts we
perform Monte Carlo simulations by randomly undersampling the 20
cluster fields having more than 50 galaxies and only one peak in the
velocity distribution.  Since our random undersampling does not
consider other parameters like the proximity to the cluster center and
the galaxy color, the following results should be considered as
conservative.

For each of the 20 cluster fields we perform 500 random simulations
extracting ten galaxies each time (the lowest limit in our
sample). Then we apply the whole procedure of member selection,
i.e. the detection of a significant peak in the velocity distribution
via the adaptive kernel method, the application of the fixed gap to
the remaining galaxies, and finally the rejection of ``active''
galaxies.  We use a fixed gap of $1250$ \ks which seemed appropriate
when a small number of galaxies is considered (cf. \S~3.1).
Only a fraction ($\sim 20\%$) of the simulated clusters survive to the
procedure of the rejection of interlopers. These clusters
contain typically 5--6 members to compute the velocity
dispersion, $\sigma_{v,i}$, and the associate statistical error,
$\Delta \sigma_{v,i}$.  For each of the 20 clusters we compute the
median value, $<\sigma_{v,i}>_{ran}$, and the s.d., $\sigma_{ran}
(\sigma_{v,i}), $ of the velocity dispersions of the corresponding
simulated clusters, as well as the median statistical error
$<\Delta\sigma_{v,i}>_{ran}$.

We verify the robustness of our estimates of velocity
dispersions by computing the median value and $90\%$ c.l. errors of
$\sigma_v/<\sigma_v>_{ran}$ within the sample of 20 clusters: we obtain
$<(\sigma_v/<\sigma_v>_{ran})>_{20}=1.01^{+0.08}_{-0.09}$ (fixed gaps
of 1000 or 1500 \ks give values of
$<(\sigma_v/<\sigma_v>_{ran})>_{20}=1.10$ or 0.97, again
consistent with unity).

These simulations also allow us to estimate the global error on the
estimate of the velocity dispersion, i.e. the error associated to the
procedure of member selection in addition to the statistical error
connected to the selected members.  On the whole sample of 20
clusters, the median value of the s.d. of velocity dispersions of
simulated clusters is very high,
$<\sigma_{ran}(\sigma_{v,i})>_{20}\sim 850$\kss, much larger than the
corresponding statistical error,
$<(<\Delta\sigma_{v,i}>_{ran})>_{20}\sim 350$ \kss.  Therefore, in the
case of a very small number of available redshifts, the error
associated to the member selection procedure can be more important
than the statistical one.  However, this kind of error rapidly
decreases as the amount of available data increases.  For instance, we
obtain a global error of $\sim 500$\ks vs. a statistical error of
$300$\ks for simulated clusters of 15 galaxies (on average 6--7
members) and the two error estimates become comparable for those
clusters containing at least ten members.

As for the poor spatial extension, as suggested by the VDPs of
Figure~1, the effect of individual clusters may be large.  To be more
quantitative, we consider the 11 clusters sampled out to $R_{vir}$ and
the corresponding velocity dispersions within $R_{vir}/2$: the
estimate of velocity dispersion varies by $\sim 25\%$ for three
clusters.  When considering the velocity dispersions within
$R_{vir}/4$ the variation is more than $25\%$ for six clusters and
reaches $65\%$ for one cluster.  Unfortunately, since the shape of
VDPs range different behaviors, the effect cannot be predicted for
individual clusters, although strongly decreasing/increasing profiles
in the external sampled regions suggest that more correct estimates of
velocity dispersions would need data over larger field of view (cf.,
e.g. AS506, CL0017-20, CL0054-27, and 3C295).

Another effect concerns the sampling within non circular
apertures. Since in our sample the elongation of the sampled region is
not extreme ($R_{max,c}/R_{max}\sim 0.6$), we expect that this effect
is smaller than the previous one.  One can quantify the effect by
increasing the weight for external cluster regions in the standard
estimate of velocity dispersion
$\sigma_{vie}^2=[\sigma_{vi}^2(N_i-1)+\sigma_{ve}^2(N_e-1)]/(N_i+N_e-1)$,
where $\sigma_{vi}$ and $\sigma_{ve}$ are the velocity dispersions as
computed on the internal and external cluster regions containing $N_i$
and $N_e$ galaxies, respectively. Possible undersampling in the
external regions can be corrected for by artificially increasing
$N_e$.  We compute $\sigma_{vi}$ and $\sigma_{ve}$ inside and outside
$R_{max,c}$ for the 35 clusters with at least 10 galaxies: a
reasonable increase of the weight (by a factor four) for the external
region leads to variations for individual clusters of the order of
$\lesssim 7\%$.

Finally, we check the effect of changing the classification of
``active'' galaxies. We consider the 22 clusters where authors
classified also galaxies with ongoing moderate star formation,
possibly spiral--like galaxies: for the data by Dressler et al. (1999)
we consider ``e(c)'' galaxies (with moderate absorption plus emission,
spiral--like); for the CNOC clusters we consider galaxies classified
with ``4'' (Sbc), and for the clusters by Couch and collaborators we
consider ``Sp'' galaxies (spiral--like, with spectra and color
properties of normal nearby spiral galaxies).  For these 22 clusters
we compare the velocity dispersions as computed in \S~4.1, $\sigma_v$,
with those computed rejecting also spiral--like galaxies as defined
above, $\sigma_{v-S}$. This different definition of ``active''
galaxies does not affect the estimate of velocity dispersions.  We
find no difference between the cumulative distributions of
$\sigma_{v}$ and $\sigma_{v-S}$ (using both the Kolmogorov--Smirnov
and the Wilcoxon tests, e.g. Press et al. 1992), the median value of
$\sigma_v/\sigma_{v-S}$ being consistent with unity.  Moreover, as for
individual clusters, $\sigma_{v}$ and $\sigma_{v-S}$ never differ by
more than $10\%$.

\section{``ACTIVE'' AND  NON ``ACTIVE'' GALAXIES}

There is evidence that the spatial distribution and kinematics of
late--type galaxies (or blue galaxies or ELGs) are different from
those of early--type galaxies (or red galaxies or NELGs); these
differences lead to higher estimates of internal velocity dispersion
and mass for clusters (e.g., Moss \& Dickens 1977; Biviano et
al. 1997; Mohr et al. 1996; Carlberg et al. 1997a; Koranyi \& Geller
2000).  Biviano et al. (1997) suggested that the dynamical state of
the ELGs, which are often spirals of late types or irregulars,
reflects the phase of galaxy infall rather than the virialized
condition in the relaxed cluster core.  Carlberg et al. (1997a)
suggested that, although differing in their distributions, both blue
and red galaxies are in dynamical equilibrium with clusters.

We check if the populations of ``active'' and non ``active'' galaxies,
AGs and NAGs, really differ in their kinematics.  Since we often
classify AGs on the base of the presence of emission lines in their
spectra (but see Couch et al. 1998 for the use of colors, too) this
classification roughly corresponds to the one between ELGs and NELGs.

\includegraphics{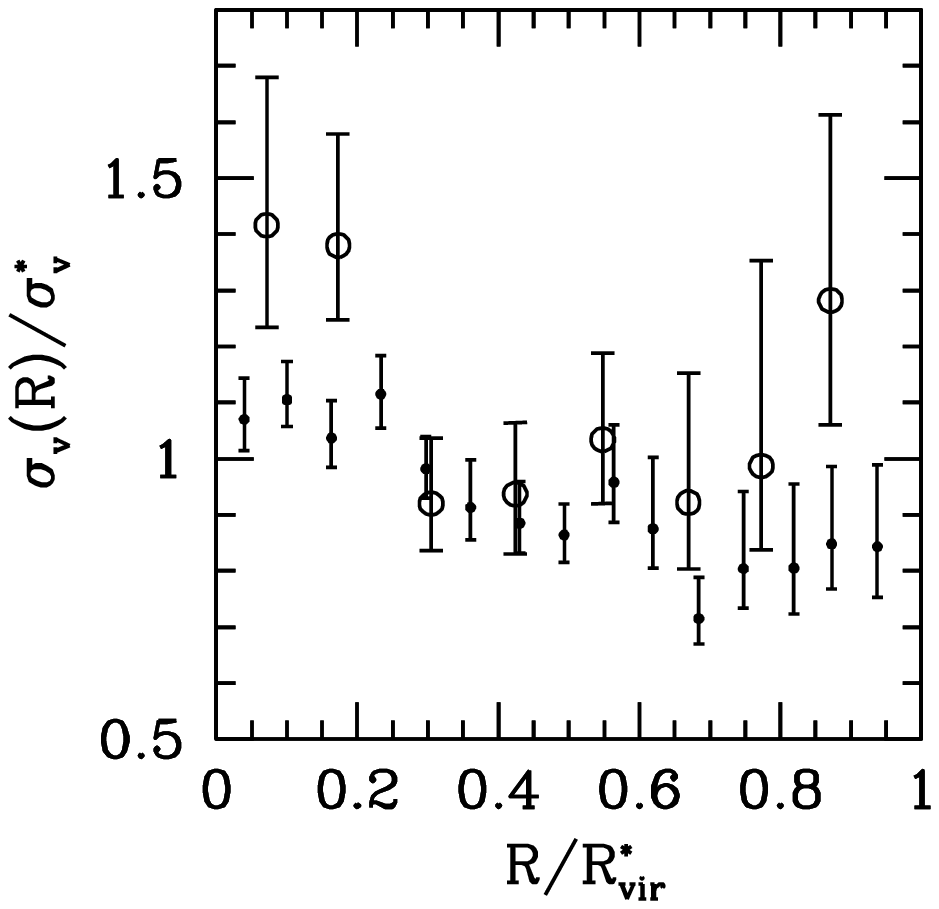}
$\ \ \ \ \ \ $\\
\vspace{7truecm}
$\ \ \ $\\
{\small\parindent=3.5mm {Fig.}~3.---
The (normalized) line--of--sight velocity
dispersion, $\sigma_v(R)$, as a function of the (normalized) projected
distance from the cluster center.  The points represent data combined
from all clusters and binned in equispatial intervals.  Open and
filled circles are obtained using the ``active'' galaxies, AGs, and
the galaxies without strong signs of activity, NAGs, respectively.  The
normalizing quantities are computed combining both the AGs and NAGs.  We
give the robust estimates of velocity dispersion and the respective
bootstrap errors.
}
\vspace{5mm}

We consider the 43 out of 51 clusters for which spectral information
is available, each cluster containing $N_{NAG}(=N_m)$ NAGs and
$N_{AG}(=N_g-N_m)$ AGs, cf. Table~2.  Figure~3 shows that the
$\sigma_v$ profile of the AGs is generally higher than the profile of
the NAGs, where the profiles are obtained combining together galaxies
of all 43 clusters and normalizing to the values of $\sigma_v$ and
$R_{vir}$ obtained for clusters before the rejection of the AGs
(otherwise the difference would be also larger). A two-dimensional
Kolmogorov--Smirnov test (Fasano \& Franceschini 1987) applied to the
normalized velocities and distances found a difference larger than
$>99\%$ between the two galaxy populations.

Only for 19 of the 43 clusters there are enough galaxies ($N_{AG}\ge5$
and $N_{NAG}\ge5$) to compute and compare the respective AG and NAG
$\sigma_v$.  Figure~4 shows the comparison between cumulative
distributions of $\sigma_v$ as computed considering the AGs or
NAGs. The AG $\sigma_v$--distribution shows a tail at high $\sigma_v$,
which, however, is not significant according to a Kolmogorov--Smirnov
test, and only slightly significant at the $93\%$ c.l.  according to
the Wilcoxon test (e.g., Press et al. 1992).  To test for different
means and dispersions of the AG and NAG populations in each individual
cluster, we apply the standard means-test and F-test (e.g., Press et
al. 1992).  We find evidence of a difference more significant than
$>95\%$ only for A851 ($\sigma_{AG}=1761$ \ks $\ne\sigma_{NAG}=1067$
\ks at the $\sim98\%$ c.l.) and for CL1602+4304 ($<V>_{AG}=268653$ \ks
$\ne<V>_{NAG}= 270349$ \ks at the $\sim96\%$ c.l.).

\includegraphics{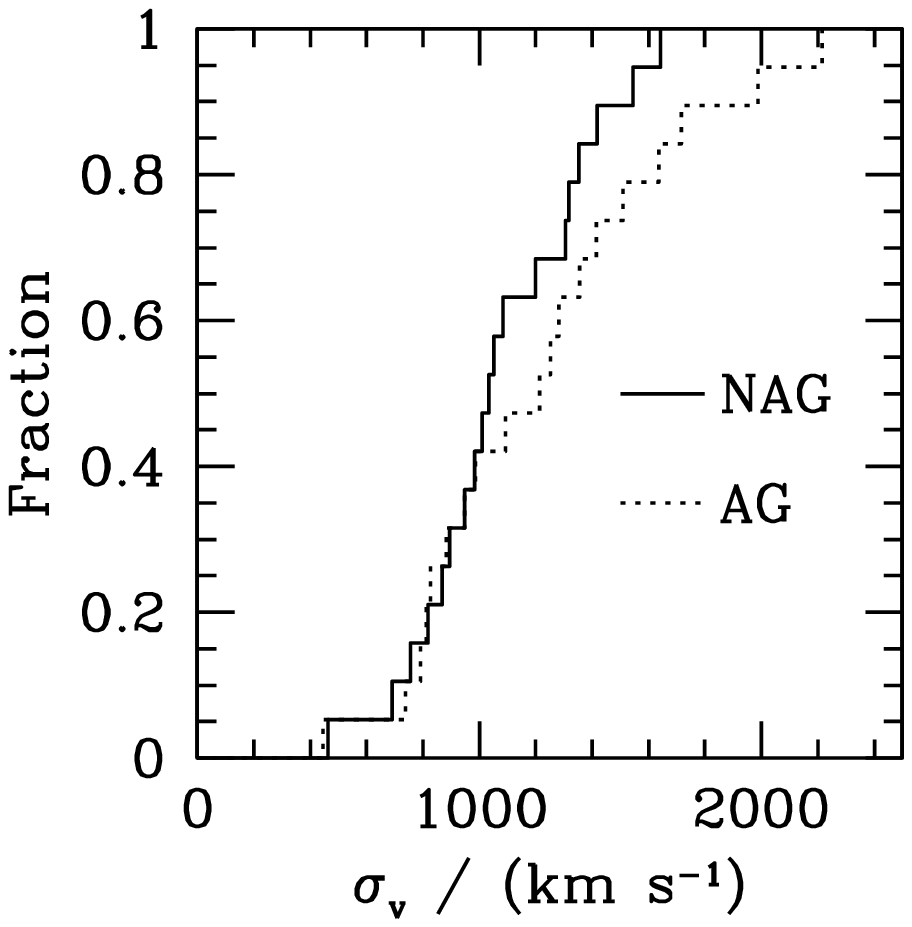}
$\ \ \ \ \ \ $\\
\vspace{7truecm}
$\ \ \ $\\
{\small\parindent=3.5mm {Fig.}~4.---
The cumulative distributions of line--of--sight velocity dispersion
computed using the 
``active'' galaxies, AGs, and the galaxies without strong signs of
activity, NAGs, as indicated by dotted and solid lines, respectively.
}
\vspace{5mm}

The evidence of difference between the two populations is in agreement
with previous findings for nearby and distant clusters (e.g., Biviano
et al. 1997; Dressler et al. 1999; Mohr et al. 1996). A really
quantitative comparison of the effect is complicated by the
differences in the definition of ``active'' galaxies.  By using ENACS
clusters, Biviano et al. (1997) found that $\sigma_v$ of ELGs is, on
average, $20\%$ larger than that of NELGs, and we find
$<\sigma_{AG}/\sigma_{NAG}>=1.12\pm0.07$.  As for the combined
velocity dispersion profile (cf. Figure~3), our result in central
cluster regions ($R\lesssim 0.1$ \h) is similar to that of of Biviano
et al (1997) giving higher central $\sigma_v$ for AGs (and ELGs) than
for NAGs (and NELGs), i.e. $\sim 1.4 \pm0.2$) vs. $\sim 1.1\pm 0.05$
for normalized $\sigma_v$. Our last point of the velocity dispersion
profile is instead very high, but we suspect that it could be due to
the loss of efficiency in rejecting interlopers in very poorly sampled
external regions of distant clusters.

\section{COMPARISON WITH RESULTS \\
FROM  X-RAY AND LENSING DATA}

We collect X-ray luminosities, in general bolometric ones,
$L_{bol,X}$, and temperature, $T_X$, for 38 and 22 clusters,
respectively.  For A223, A521, CL0054-27, CL0412-65, CL1604+4304, and
the four clusters by Bower et al. (1997) we obtain the bolometric
luminosities by multiplying the original band luminosities by a
temperature-dependent bolometric correction factor. This factor is
computed under the assumptions of pure bremsstrahlung intracluster
medium emission and a power--law approximation for the Gaunt factor.
For the correction we use the temperatures estimated from $\sigma_v$
in the hypothesis of density energy equipartition between hot gas and
galaxies, i.e.  $\beta_{spec}=\sigma_v^2/(kT/\mu m_p)=1$, where
$\mu=0.58$ is the mean molecular weight and $m_p$ the proton mass.
For the four clusters by Bower et al. (1997), which have very few
galaxies with measured redshift, we use the $L_X-\sigma_v$ relation given
by the authors for nearby clusters. For these four clusters we expect
$\sigma_v\sim$ 600 \kss, i.e. $T\sim2$ keV.

In Table~4 we list the collected values for $L_{bol,X}$, and $T_X$
with the corresponding reference sources (Cols.~2--5), and the value
of $\beta_{spec}$ (Col.~6). The errors on $\beta_{spec}$ 
take into
account errors on both $\sigma_v$ and $T_X$.

Figure~5 shows the $L_{X,bol}$--$\sigma_v$ relation compared to that
found by Borgani et al. (1999) for nearby clusters.  
Excluding the leftmost point
(J2175.15TR), the resulting bisecting linear regression is
\begin{equation}
log(L_{bol,X}/10^{44}erg\,s^{-1})=4.4^{+1.8}_{-1.0}\,log(\sigma_v/km \,s^{-1})-12.6^{+3.0}_{-5.4}\,,
\end{equation}
\noindent where errors come from the difference with respect to the
direct and the inverse linear regression (Isobe et al. 1990, OLS
methods).  Our $L_{bol,X}$--$\sigma_v$ relation is consistent with
that of nearby clusters (e.g., White et al. 1997; Borgani et al. 1999;
Wu, Xue, \& Fang 1999).  As for the point excluded, note that our
analysis of J2175.15TR is based only on 19 galaxies, and the estimate
of $\sigma_v$ is recovered from only eight member galaxies (with an
error larger than $100\%$).

All clusters for which $\sigma_v$ should be better interpreted as an
upper limit to the true estimate (AS506, CL0054-27, 3C295) lie on the
right--upper corner of the plot. Excluding these points we fit a
consistent relation, i.e.
$log(L_{bol,X}/10^{44}erg\,s^{-1})=4.7^{+1.9}_{-1.1}\,log(\sigma_v/km\,s^{-1})-13.5^{+3.1}_{-5.5}$.

Figure~6 shows the $\sigma_v$--$T_X$ relation compared to that of
nearby clusters, as reported by G98.  As for distant 

\end{multicols}
\hspace{-9mm}
\begin{minipage}{20cm}
\renewcommand{\arraystretch}{1.2}
\renewcommand{\tabcolsep}{1.2mm}
\begin{center}
TABLE 4\\
{\sc Comparison with X--ray Properties \\}
\footnotesize 
\input{tab4}
\end{center}
\input{comm_tab4}
\end{minipage}
\begin{multicols}{2}

\noindent clusters, the
data have a too small dynamical range to attempt a linear fit: the
visual inspection of Figure~6 suggests no difference with nearby
clusters in agreement with the result of by Mushotzky \& Scharf (1997)
and Wu et al. (1999).  We obtain $\beta_{spec}=0.88^{+0.14}_{-0.17}$
(median value with errors at the $90\%$ c.l.), in agreement with the
value of $\beta_{spec}=0.88\pm0.04$ for nearby clusters (cf. G98).
Moreover, we find no correlation between $\beta_{spec}$ and redshift
(cf. also Wu et al. 1999).

Under the assumption that the hot diffuse gas is in hydrostatic and
isothermal equilibrium with the underlying gravitational potentials of
clusters, one can obtain the X--ray cluster masses provided that the
gas temperature and radial profile of gas distribution are known
(e.g. Wu et al. 1998).  The availability of $T_X$ allow us to compute
the mass within $R_{vir}$ for 22 clusters according to
$M_{X}=(3\beta_{fit,gas}kT\cdot R_{vir})/(G \mu m_p) \cdot
(R_{vir}/R_x)^2/[1+(R_{vir}/R_x)^2]$, 
where we adopt the gas
distribution given by the $\beta$--model with typical parameters
(slope $\beta_{fit,gas} =2/3$ and core radius $R_x=0.125$ \h,
e.g. Jones \& Forman 1992).  
We find mass values consistent with our
optical virial estimates, i.e.  $M/M_{X}=1.02\,(0.86$--$1.32)$ for the
median value and the range at the $90\%$ c.l..

\includegraphics{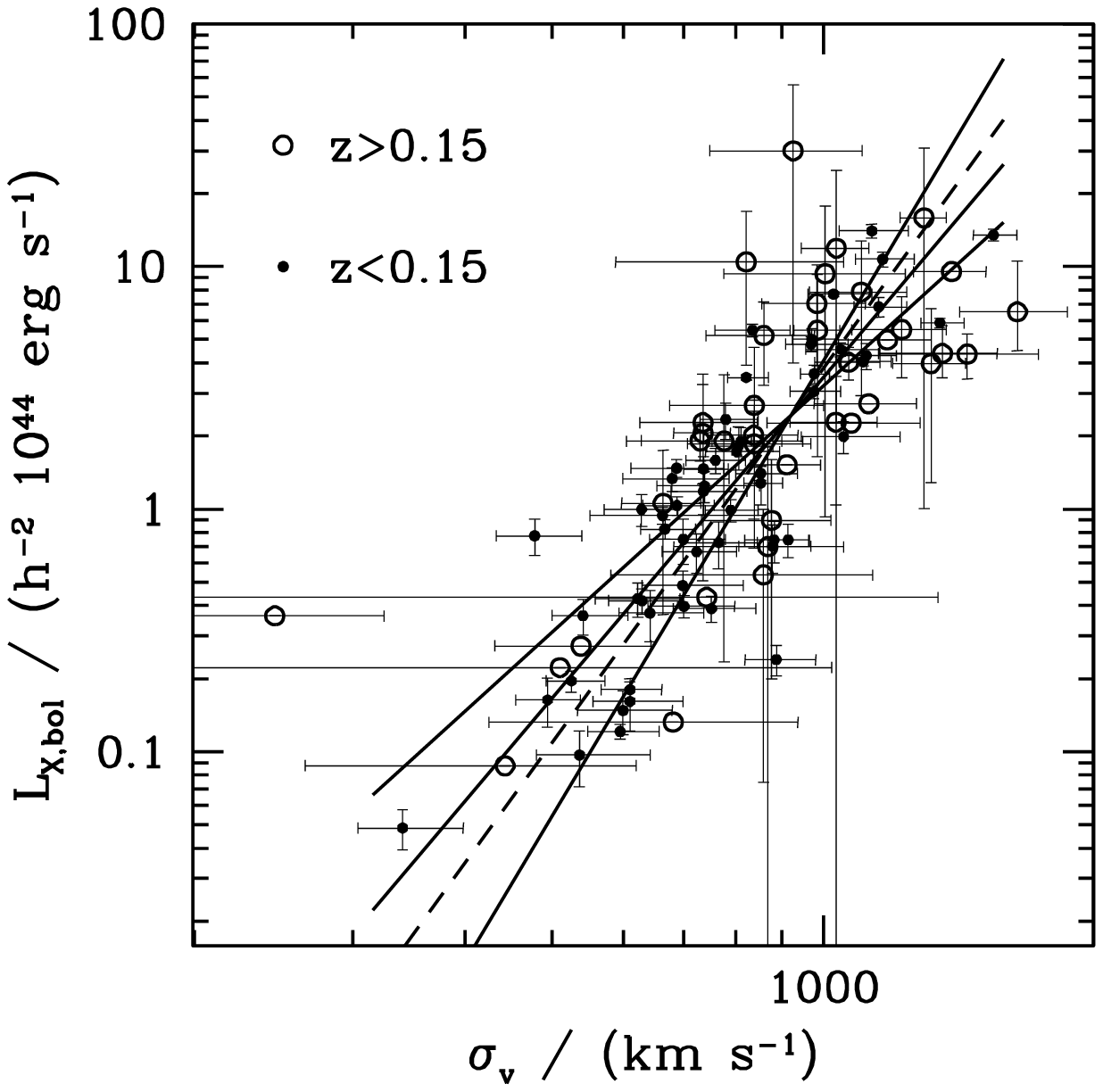}
$\ \ \ \ \ \ $\\
\vspace{8.5truecm}
$\ \ \ $\\
{\small\parindent=3.5mm {Fig.}~5.---
$L_{X,bol}$--$\sigma_v$ relation for distant
(open circles) and nearby clusters (filled circles).  For the nearby
clusters we show results as reported by Borgani et al. (1999), all
having $\sigma_v$ estimated at least with 30 galaxy redshifts (Girardi
et al. 1998b) and also belonging to the X--ray Brightest Abell--like
Cluster survey (Ebeling et al. 1996).  The error bands at the $68\%$
c.l. are shown: errors on $L_{X,bol}$ are not available for a few
distant clusters.  The three solid lines are direct, inverse, and
bisecting linear regression for the distant clusters (obtained
rejecting the point on the left).  The dashed line is the bisecting
linear regression for the nearby clusters as computed by Borgani et
al. (1999).
}
\vspace{5mm}

As for gravitational lensing masses, we resort to estimates found in
the literature. We collect projected estimates from weak gravitational
lensing analysis, $M_L$, for 18 clusters.  In order to compare our
optical virial masses to $M_L$, we project and rescale our masses $M$
within the corresponding radius using the fitted galaxy spatial distribution.
In Table~5 we list the reference sources (Col.~2) from where we take
$R_L$ and $M_L$ (Cols.~3 and 4); the corresponding optical virial
projected mass, $M_{opt,L}$ (Col.~5); and the respective ratio
(Col.~6).  We obtain $M_{opt,L}/M_L=1.30(0.63$--$2.13)$ (median value
and range at the $90\%$ c.l.).  Moreover, we do not find any
correlation between $M/M_X$ or $M_{opt,L}/M_L$ and redshift.

Our finding are in agreement with other recent studies which find, on
average, no evidence of discrepancy between different mass estimates
as computed within large radii, thus suggesting that distant clusters
are nor far from global dynamical equilibrium (e.g., Allen 1998; Wu et
al. 1998; Lewis et al. 1999).  Note that we avoid to consider mass
determination in very central cluster regions since our analysis of
cluster members give poor constrains on mass distribution on these
scales.  Indeed, the assumption of dynamical equilibrium seem to break
down in the very central regions as suggested by comparisons with
strong lensing mass estimates (e.g., Allen 1998; Lewis et al. 1999; Wu
et al. 1998).

\includegraphics{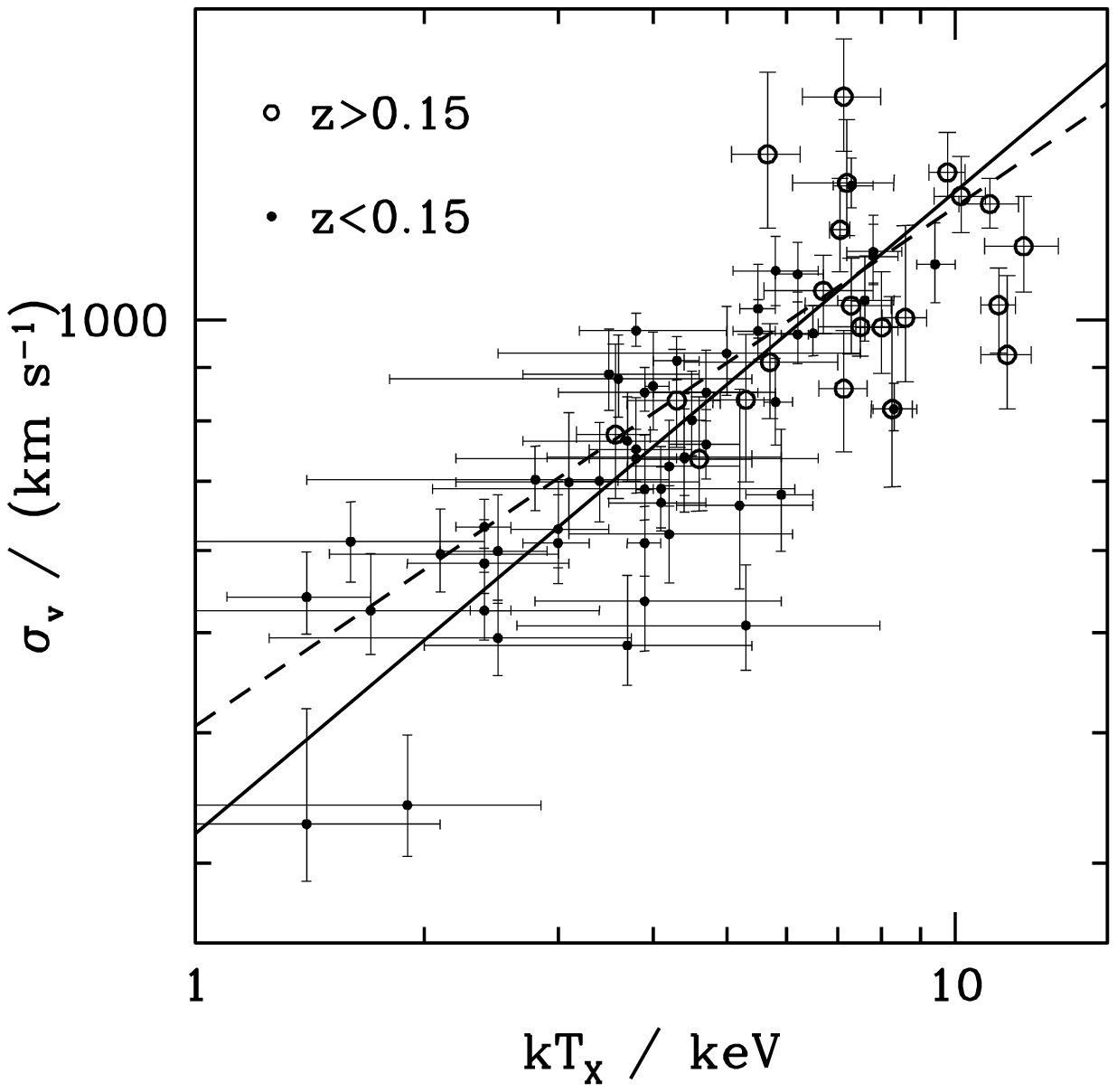}
$\ \ \ \ \ \ $\\
\vspace{8.5truecm}
$\ \ \ $\\
{\small\parindent=3.5mm {Fig.}~6.---
$\sigma_v$--$T_X$ relation for distant (open
circles) and nearby clusters (filled circles).  For the nearby
clusters we show results as reported by Girardi et al. (1998b), all
having $\sigma_v$ estimated at least with 30 galaxy redshifts, and
$T_X$ taken from David et al. (1993) and from White et al. (1997).
The error bands at the $68\%$ c.l. are shown: when authors give only
$90\%$ c.l. errors on $T_X$, we apply a reduction by a factor of 1.6.
The solid line is the bisecting linear regression for the nearby
clusters as computed by Girardi et al. (1998b).  The dashed line represents
the model with the equipartition of energy per unit mass between gas
and galaxy components ($\beta_{spec}=1$).
}
\vspace{5mm}

\section{SUMMARY AND CONCLUSIONS}

In order to properly analyze the possible dynamical evolution of
galaxy clusters we apply the same procedures already applied on a
sample of nearby clusters (170 clusters at $z<0.15$ from ENACS and
other literature, Girardi et al. 1998b, cf. also Fadda et al. 1996) to
a corresponding sample of 51 distant clusters ($0.15\lesssim z
\lesssim 0.9$, $<z>\sim0.3$).  Each cluster has at least 10 galaxies
with available redshift in the literature.  Three cluster fields show
two overlapping peaks in their velocity distribution and large
uncertainties in their dynamics. Out of other cluster fields, 45
fields show only one peak in the velocity distribution and three
fields show two separable peaks, for a total of 51 well--defined
cluster systems.  These 51 systems are those

\end{multicols}
\hspace{-9mm}
\begin{minipage}{20cm}
\renewcommand{\arraystretch}{1.2}
\renewcommand{\tabcolsep}{1.2mm}
\begin{center}
TABLE 5\\
{\sc Comparison with Masses from Weak Gravitational Lensing \\}
\footnotesize 
\input{tab5}
\end{center}
\input{comm_tab5}
\end{minipage}
\begin{multicols}{2}

\noindent used in the comparison
with nearby clusters (i.e., 160 well--defined systems, cf. \S~3 of
Girardi et al. 1998b).

We select member galaxies, analyze the velocity dispersion profiles,
and evaluate in a homogeneous way cluster velocity dispersions and
virial masses.

As a main general result, we do not find any significant evidence for
dynamical evolution of galaxy clusters. More in detail, our results
can be summarized as follows.

\begin{itemize}

\item The galaxy spatial distribution is similar to that of nearby
clusters, i.e. the fit to a King--like profile gives a two-dimensional
slope of $\alpha=0.7$ and a very small core radius of $R_c=0.05$
\hh. Note that we do not want to really asses the existence of a core,
or to state that the King--modified profile is better than other forms
for galaxy density profiles; the King--modified profile is used for a
consistent comparison with nearby clusters.  We refer to Girardi et
al. (1998b, \S~8) for other relevant analyses and discussions.

\item For those clusters with good enough data, 
the integrated velocity dispersion
profiles show a behavior similar to those of
nearby clusters: they are strongly increasing or decreasing
 in the central cluster regions, but always flattening out in
the external regions, thus suggesting that 
large--scale dynamics is not 
affected by velocity anisotropies.

\item The average velocity dispersion profile can be explained by a
model with isotropic orbits, which well describe also nearby clusters.
Possible evidences for more radial orbits are not statistically
significant.

\item There is no evidence of evolution in both
$L_{bol,X}$--$\sigma_v$ and $\sigma_v$--$T_X$ relations, thus in
agreement with previous results (Mushotzky \& Scharf 1997; Borgani et
al. 1999).

\end{itemize}

Moreover, within the large scatter of present data, we find, on
average, no significant evidence of discrepancies between our virial
mass estimates and those from X--ray and gravitational lensing data,
thus suggesting that distant clusters are not far from global
dynamical equilibrium (cf. also Allen 1998; Lewis et al. 1999; Wu et
al. 1998).

We conclude that the typical redshift of cluster formation is higher
than that of our sample in agreement with previous suggestions (e.g.,
Schindler 1999; Mushotzky 2000).  In particular,
we agree with preliminary results by Adami et al. (1999), who applied
the same techniques used for the nearby ENACS clusters on 15 distant
clusters, ($<z>\sim0.4$) from the Palomar Distant Cluster Survey
(Postman et al. 1996).

Although some clusters at very high redshift, e.g. $z>0.8$, are
already known (e.g., Gioia et al.  1999; Rosati et al. 1999), the
construction of a large cluster sample useful for studying internal
dynamics will require a strong observational effort. Note that,
already in the construction of the cluster sample analyzed here, we
relax the requirements applied to the sample of nearby clusters by
Girardi et al. (1998b), i.e. the distant clusters are more poorly
sampled.  Throughout the presentation of our analyses we stress how
both the poor number of galaxies and the small spatial extension of
some clusters can affect the robustness of their resulting properties.
In particular, Monte Carlo simulations, which take into account the
whole membership procedure, show that the estimate of velocity
dispersion is, on average, well recovered also in the case of very
poor sampling (only 10 galaxies in the cluster field giving 5--6
members), but that the global error associated to the individual
clusters should be a factor $\sim 2.5$ larger than the pure
statistical error.  Also the small spatial extension could lead to
large over/underestimates of velocity dispersion of individual
clusters: we evaluate that variations of $25\%$ are quite common when
clusters are sampled only out to half of the virialization region.

\acknowledgments

We thank the anonymous referee for useful suggestions and
comments.  We also thank Andrea Biviano and Stefano Borgani for
interesting discussions, and Massimo Ramella for providing us
redshifts and positions of galaxies for 1E0657-56.  Special thanks to
Alenka Devetak for her help in initial phase of this project.  This
research has made use of the NASA/IPAC Extragalactic Database (NED)
which is operated by the Jet Propulsion Laboratory, California
Institute of Technology, under contract with the National Aeronautics
and Space Administration.  This work has been partially supported by
the Italian Ministry of University, Scientific Technological Research
(MURST), by the Italian Space Agency (ASI), and by the Italian
Research Council (CNR-GNA).

$\ $

$\ $

\appendix
\section{RESULTS FOR \\
MULTIPEAKED CLUSTERS}

Here we shortly present the results of our analysis for the three
clusters with uncertain dynamics, i.e. with two peaks in the velocity
distribution which are not clearly separable. We consider both the
system composed by the two peaks together and each peak individually.
In Figure~7 we plot the velocity-space galaxy density for each
cluster, as provided by the adaptive--kernel reconstruction method,
and the integrated velocity dispersion profile VDP for each possible
system.  Table~6 summarizes the results of the analysis of the
internal dynamics. Note the strong variation in $\sigma_v$ and mass
when considering the two peaks together or not.  Some comments on
individual clusters follow.

\includegraphics{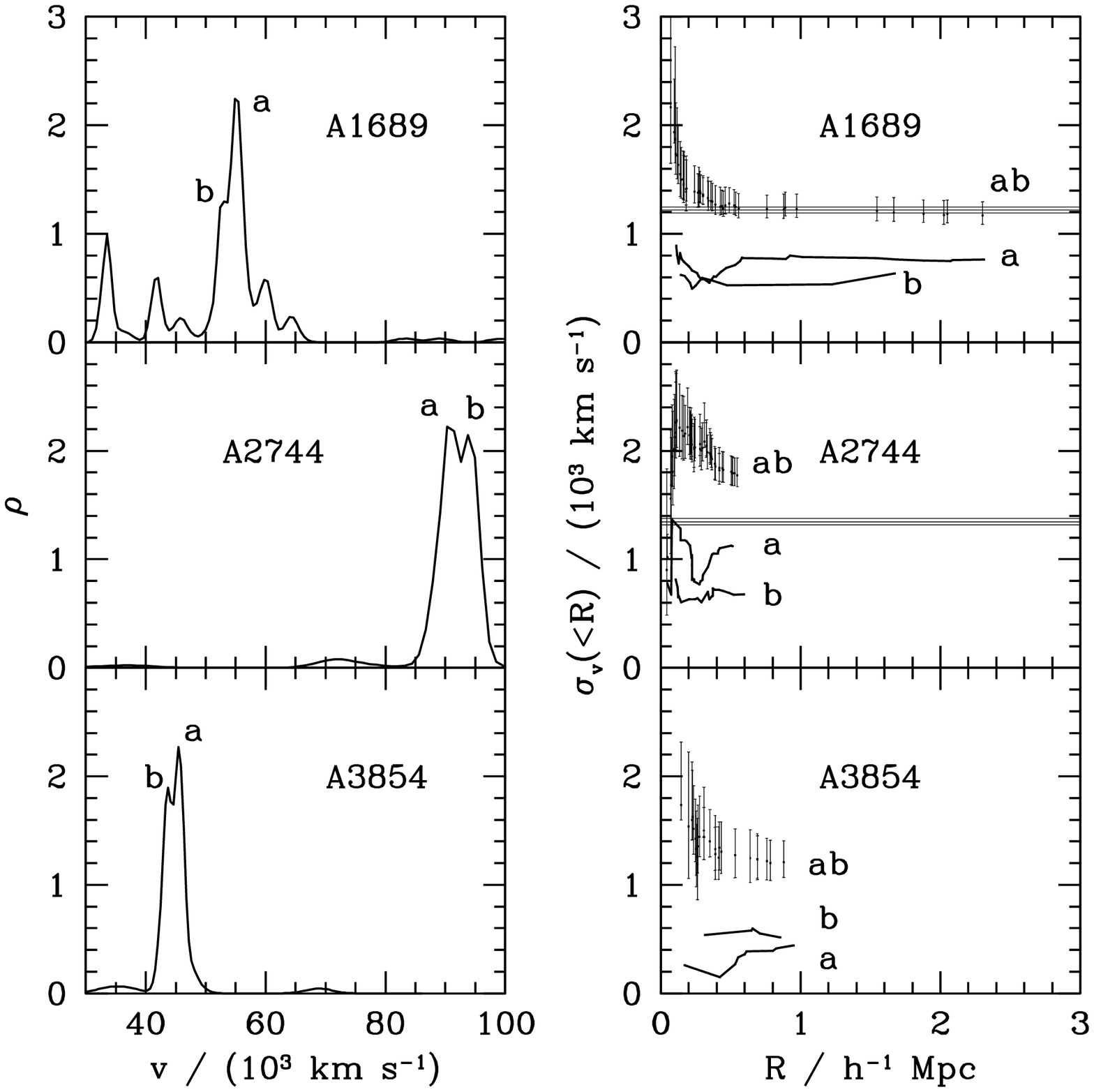}
$\ \ \ \ \ \ $\\
\vspace{9.5truecm}
$\ \ \ $\\
{\small\parindent=3.5mm {Fig.}~7.---
For each of the three clusters we give the
relative velocity-space galaxy density, as provided by the adaptive
kernel reconstruction method (left panels) and integrated
line--of--sight velocity dispersion profiles $\sigma_v(<R)$ for each
of the considered systems (right panels).  For the velocity dispersion
profiles we plot bootstrap errors only in the case of the system with
joined peaks.  The horizontal lines in the right panels represent
X--ray temperature with the respective errors taken from Wu et
al. (1999) and transformed in $\sigma_v$ imposing $\beta_{spec}=1$.
}
\vspace{5mm}

\end{multicols}
\hspace{-9mm}
\begin{minipage}{20cm}
\renewcommand{\arraystretch}{1.2}
\renewcommand{\tabcolsep}{1.2mm}
\begin{center}
TABLE 6\\
{\sc Clusters with Uncertain Dynamics \\}
\footnotesize 
\input{tab6}
\end{center}
\end{minipage}
\begin{multicols}{2}

\noindent{\bf A1689} Teague, Carter, \& Gray (1990) computed a value
of $\sigma_v=1989$ \kss.  As for the analysis of the cluster members,
the two peaks in the velocity distribution were already pointed out by
Girardi et al. (1996; 1997b) using the same adaptive kernel method.
By using a multi--scale analysis which couples kinematical estimators
with the wavelet transform, Girardi et al. (1997b) found the presence
of two dynamically relevant structures, but with a smaller $\sigma_v$
and mass with respect to the two systems, ``a'' and ``b'', analyzed
here.  Moreover, A1689 is well know for a strong discrepancy between
mass from X-ray and strong gravitational lensing analyses
(e.g. Miralda-Escud\`e \& Babul 1995) which could be due to its
complex structure.  Also the very recent weak lensing analysis of
Taylor et al. (1998) suggests the model of a double cluster aligned
along the line of sight in order to explain discrepancies between
optical and X--ray results. These results and the fact that A1689
appears well aligned along the line of sight with other structures
(three foreground groups, Teague et al.  1990) suggests the presence
of a large structure filament well aligned along the line of sight.

\noindent{\bf A2744} Couch \& Sharples (1987) computed a value of
$\sigma_v=1947^{+292}_{-201}$ \kss.  A strong suggestion for the
dynamical activity comes from the recent study by Allen (1998): among
the 13 clusters analyzed, A2744 shows the strongest discrepancy
between mass from X--ray and gravitational lensing analyses.

\noindent{\bf A3854} Colless \& Hewett (1987) listed a value of
$\sigma_v=1180^{+202}_{-143}$ \kss.

\end{multicols}

\end{document}

%% file: tab1.tex
%
%
\begin{tabular}{llrc}
\hline \hline
\multicolumn{1}{c}{Cluster Name}
&\multicolumn{1}{c}{Other Names}
&\multicolumn{1}{c}{$N$}
&\multicolumn{1}{c}{References}
\\
\multicolumn{1}{c}{(1)}
&\multicolumn{1}{c}{(2)}
&\multicolumn{1}{c}{(3)}
&\multicolumn{1}{c}{(4)}
\\
\hline
A115$^a$     &ZwCl0053.4+2604& 28&1\\
A140$^a$     &EDCC 520& 11&2\\
A222& &33& 3\\
A223&&28& 3\\
A370     && 58&4\\
A520&MS0451.5+0250&27& 3\\
A521     && 49&5\\
A665     &ZwCl0826.1+6554& 41&6\\
A851     &Cl 0939+47&137&4,7\\
A1300    && 95&8\\
A1689    &&130&9\\
A2218    && 53&10\\
A2390    &RXJ2153.6+1741&325&11\\
A2744    &AC118& 76&12,13\\
A3639    && 14&14\\
A3854    &C52& 41&15\\
A3888    &CL22315-3800& 98&9\\
A3889    && 26&16\\
AS506\tablenotemark{a}    &CL0500-24& 29&17\\
AS910    &AC103& 88&12,13\\
AS1077   &AC114&103&13,18\\
CL0017-20\tablenotemark{a}    && 26&17\\
CLJ0023+0423  &GHO0021+0406 &107&19,20\\
CL0024+16    &ZwCl0024.0+1652 &134&4,7\\
CL0053-37    && 22&16\\
CL0054-27    &J1888.16CL& 25&4\\
CL0303+17    &GHO0303+1706 & 84&4\\
CL0412-65    && 24&4\\
CL0949+44    &GHO0949+4408 & 33&7\\
CL1447+26    && 29&4\\
CL1601+42    &GHO1601+4259 &101&4,7\\
CLJ1604+4304  &GHO1602+4312 & 95&19,20\\
F1637.23TL    && 19&21\\
F1652.20CR    && 20&21\\
J2175.15TR    && 19&21\\
J2175.23C     && 19&21\\
MS0015.9+1609&CL0016+1609 HSTJ001831+16207&111&22\\
MS0302.7+1658&Cl0302+1658& 96&23\\
MS0302.5+1717&Cl0302+1717& 43&24\\
MS0440.5+0204\tablenotemark{a}&& 56&25\\
MS0451.6-0305&&113&22\\
MS1008.1-1224&&109&26\\
MS1054.4-0321\tablenotemark{a}&& 32&27\\
MS1224.7+2007&& 54&28\\
MS1358.4+6245&ZwCl1358.1+6245&281&26\\
MS1512.4+3647&&282&28\\
MS1621.5+2640&&262&23\\
RXJ1716+67    && 37&29\\
1E0657-56\tablenotemark{a}&RASS1 069 & 32& 30\\
3C206\tablenotemark{a}&&15&31\\
3C295 $\ \ \ \ \ \ \ \ \ \ \ \ \ \ \ \ \ \ \ \ \ \ $ & $\ \ \ \ \ \ \ \ \ \ \ \ \ \ \ \ \ \ \ \ \ \ \ \ \ \ \ \ \ \ \ \ \ \ $& 38&4,7\\
\hline
\end{tabular}

%% file: comm_tab1.tex
{\footnotesize\parindent=3mm 
$^a$~Clusters for which spectral/morphological information is not 
available.\\ 
(1) Zabludoff et al. 1990; (2) Collins et al. 1995; (3) Proust
et al. 2000; (4) Dressler et al. 1999; (5) Maurogordato et al. 1999;
(6) Oegerle et al. 1991; (7) Dressler \& Gunn 1992; (8) Lemonon et
al. 1997; (9) Teague et al. 1990; (10) Le Borgne et al. 1992;
(11) Yee et al. 1996a; (12) Couch et al. 1998; (13) Couch \& Sharples 
1987; (14) Garilli et al. 1991; (15) Colless \&  Hewett 1987; (16)
Cappi et al. 1998; (17) Infante et al. 1994; (18) Couch et al. 1994; 
(19) Lubin et al. 1998b; (20) Postman et al. 1998; (21) Bower et al. 1997;
(22) Ellingson et al. 1998; (23) Ellingson et al. 1997; (24) Fabricant
et al. 1994; (25) Gioia et al. 1998; (26) Yee et al. 1998; (27) Tran et
al. 1999; (28) Abraham et al. 1998; (29) Gioia et al. 1999; (30) Tucker
et al. 1998; (31) Ellingson et al. 1989.\\
Cluster names: ``A'' for the catalog of Abell et al. (1989) and,
in particular, ``AS'' for the supplementary southern clusters; ``AC'' and
``C'' for clusters used by Couch \& Newell (1984), and by Colless \&
Hewett (1987), respectively, taken from the southern extension of the
Abell catalog (Abell et al. 1989) in preparation at that time; ``EDCC''
for the Edinburgh--Durham Cluster Catalog (Lumsden et al. 1992); ``F'' and
``J'' for the catalog of Couch et al. (1991); ``GHO'' for the catalog of
Gunn et al. (1986); ``HST'' for the Hubble Space Telescope Medium Deep
Survey cluster sample of Ostrander et al. (1998); ``MS'' for the Extended
Medium Sensitivity Survey (Gioia et al. 1990); ``RASS1'' for the sample of
bright clusters of galaxies in the southern hemisphere (De Grandi et al.
1999), based on the first analysis of the ROSAT All-Sky Survey data
(RASS1); ``ZwCl'' for the catalog of Zwicky et al. (1961-1968); ``CL'' for
clusters optically detected at the given coordinates; ``RX'' for ROSAT
X--ray clusters;  1E0657-56 is a cluster detected by the imaging
proportional counter (IPC) on board the Einstein Observatory; ``3C'' for
clusters surrounding the corresponding radio source of the 3C Revised
Catalogue (Bennett 1962).

}

%% file: tab2.tex
%
%

\begin{tabular}{lrrrlcc}
\hline\hline
\multicolumn{1}{c}{Cluster Name}
&\multicolumn{1}{c}{$N_p$}
&\multicolumn{1}{c}{$N_g$}
&\multicolumn{1}{c}{$N_m$}
&\multicolumn{1}{c}{$<z>$}
&\multicolumn{2}{c}{Center}
\\
\multicolumn{1}{c}{}
&\multicolumn{1}{c}{}
&\multicolumn{1}{c}{}
&\multicolumn{1}{c}{}
&\multicolumn{1}{c}{}
&\multicolumn{1}{c}{}{R.A.(J2000)}
&\multicolumn{1}{c}{}{$\delta$(J2000)}
\\
\multicolumn{1}{c}{(1)}
&\multicolumn{1}{c}{(2)}
&\multicolumn{1}{c}{(3)}
&\multicolumn{1}{c}{(4)}
&\multicolumn{1}{c}{(5)}
&\multicolumn{2}{c}{(6)}
\\ 
\hline
 A115S &   16&   13 &  13 &  .1958 &00:55:59.90&$+$26:20:08.3\\
 A140          &   7 &   7  &   7 &  .1600 &01:04:31.55&$-$23:57:46.2\\
 A222          &   29&   27 &  26 &  .2138 &01:37:33.83&$-$12:59:21.1\\
 A223          &   18&   18 &  14 &  .2119 &01:37:56.48&$-$12:48:33.9\\
 A370          &   38&   37 &  35 &  .3744 &02:39:51.58&$-$01:34:12.4\\
 A520          &   19&   18 &  18 &  .2000 &04:54:14.42&$+$02:57:14.8\\  
 A521     &   40&   37 &  35 &  .2474 &04:54:09.13&$-$10:14:25.1\\
 A665     &   31&   26 &  25 &  .1806 &08:30:46.85&$+$65:53:52.9\\
 A851     &   67&   65 &  55 &  .4061 &09:42:58.15&$+$46:59:34.9\\
 A1300    &   62&   59 &  53 &  .3078 &11:31:57.69&$-$19:54:35.2\\
 A1689a   &   41&   38 &  38 &  .1837 &13:11:31.62&$-$01:20:58.0\\
 A1689b   &   16&   16 &  15 &  .1746 &13:11:28.58&$-$01:20:25.3\\
 A1689ab  &   57&   50 &  49 &  .1821 &13:11:30.24&$-$01:20:54.2\\
 A2218    &   45&   43 &  43 &  .1761 &16:35:51.96&$+$66:12:19.8\\
 A2390    &  243&   211&  200&  .2282 &21:53:36.80&$+$17:41:32.2\\
 A2744a   &   36&   36 &  34 &  .3014 &00:14:21.26&$-$30:23:49.3\\
 A2744b   &   27&   26 &  25 &  .3148 &00:14:20.57&$-$30:24:04.4\\
 A2744ab  &   63&   57 &  55 &  .3078 &00:14:21.16&$-$30:23:52.4\\
 A3639    &   8 &   8  &  7  &  .1480 &19:28:18.41&$-$50:54:28.6\\
 A3854a   &   21&   18 &  18 &  .1520 &22:17:53.03&$-$35:46:42.7\\
 A3854b   &   13&   13 &  9  &  .1459 &22:17:33.97&$-$35:44:47.1\\
 A3854ab  &   34&   34 &  30 &  .1506 &22:17:46.50&$-$35:45:12.3\\
 A3888    &   81&   55 &  50 &  .1508 &22:34:26.90&$-$37:43:51.0\\
 A3889a   &   7 &   7  &  7  &  .2559 &22:34:54.47&$-$30:33:50.5\\
 A3889b   &   9 &   9  &  9  &  .2495 &22:34:49.35&$-$30:32:13.1\\
 AS506         &   23&   21 &  21 &  .3201 &05:01:11.85&$-$24:25:01.5\\
 AS910         &   56&   54 &  53 &  .3076 &20:57:02.89&$-$64:40:04.7\\
 AS1077         &   85&   70 &  63 &  .3148 &22:58:47.14&$-$34:47:59.8\\
 CL0017-20    &   20&   20 &  20 &  .2717 &00:19:37.82&$-$20:26:39.1\\
 CLJ0023+0423a  &   14&   14 &  5  &  .8453 &00:23:52.69&$+$04:19:38.3\\
 CLJ0023+0423b  &   7 &   7  &  5  &  .8273 &00:23:53.82&$+$04:23:16.2\\
 CL0024+16    &  102&   95 &  73 &  .3937 &00:26:34.79&$+$17:10:04.8\\
 CL0053-37    &   20&   20 &  20 &  .1652 &00:55:59.44&$-$37:32:36.1\\
 CL0054-27    &   10&    9 &  7  &  .5604 &00:56:56.04&$-$27:40:31.9\\
 CL0303+17    &   44&   43 &  29 &  .4195 &03:06:14.13&$+$17:18:09.0\\
 CL0412-65     &   7 &   7  &  6  &  .5086 &04:12:53.39&$-$65:51:13.2\\ 
 CL0949+44a   &   15&   15 &  14 &  .3781 &09:52:57.50&$+$43:55:37.0\\
 CL0949+44b   &   9 &   9  &  8  &  .3493 &09:53:01.25&$+$43:55:22.6\\
 CL1447+26    &   16&   11 &  5  &  .3763 &14:49:28.78&$+$26:06:58.3\\
 CL1601+42    &   57&   53 &  46 &  .5403 &16:03:10.46&$+$42:45:37.2\\
 CLJ1604+4304    &   19&   14 &  8  &  .9018 &16:04:25.09&$+$43:04:11.0\\
 F1637.23TL    &   8 &   8  &  6  &  .47903 &23:59:20.68&$-$32:17:45.3\\
 F1652.20CR    &   8 &   8  &  6  &  .4102 &04:47:57.66&$-$20:37:29.6\\
 J2175.15TR    &   8 &   8  &  8  &  .3948 &03:34:20.71&$-$38:53:54.5\\
 J2175.23C     &   10&    6 &  5  &  .4058 &03:32:59.30&$-$39:06:49.7\\
 MS0015.9+1609    &   63&   50 &  42 &  .5490 &00:18:33.49&$+$16:26:02.5\\
 MS0302.7+1658    &   37&   33 &  30 &  .4248 &03:05:31.63&$+$17:10:12.0\\
 MS0302.5+1717    &   28&   26 &  24 &  .4241 &03:05:17.92&$+$17:28:34.4\\
 MS0440.5+0204    &   32&   32 &  32 &  .1969 &04:43:09.46&$+$02:10:29.5\\
 MS0451.6-0305    &   67&   46 &  40 &  .5403 &04:54:11.24&$-$03:00:45.4\\
 MS1008.1-1224    &   74&   71 &  65 &  .3070 &10:10:31.76&$-$12:40:05.4\\
 MS1054.4-0321    &   32&   32 &  32 &  .8318 &10:56:57.31&$-$03:37:44.2\\
 MS1224.7+2007    &   23&   23 &  23 &  .3253 &12:27:18.81&$+$19:50:26.7\\
 MS1358.4+6245    &  185&   141\tablenotemark{a}& 133&  .3278 &13:59:50.92&$+$62:30:49.8\\
 MS1512.4+3647    &   70&   46 &  35 &  .3711 &15:14:16.45&$+$36:34:57.7\\
 MS1621.5+2640    &  119&   106&  88 &  .4271 &16:23:34.52&$+$26:34:17.1\\
 RXJ1716+67    &   37&   32 &  19 &  .8073 &17:16:48.92&$+$67:08:21.1\\
 1E0657-56     &   14&   12 &  12 &  .2966 &06:58:37.83&$-$55:56:56.0\\
 3C206         &   7 &   7  &  7  &  .1980   &08:39:50.10&$-$12:14:32.2\\ 
 3C295         &   21&   21 &  15 &  .4591 &14:11:21.54&$+$52:11:54.6\\
\\
\hline
\end{tabular}

%% file: comm_tab2.tex
{\footnotesize\parindent=3mm
$^a$~ 
Only galaxies within 1.2 \h from the cluster
center (cf. \S~3.1).
}

%% file: tab3.tex
%
%

\begin{tabular}{lrrlrrcll}
\hline\hline
\multicolumn{1}{c}{Name}
&\multicolumn{1}{c}{$N_m$}
&\multicolumn{1}{c}{$R_{max}$}
&\multicolumn{1}{c}{$\sigma_v$}
&\multicolumn{1}{c}{$R_{vir}$}
&\multicolumn{1}{c}{$R_{PV}$}
&\multicolumn{1}{c}{$T$}
&\multicolumn{1}{c}{$M_V$}
&\multicolumn{1}{c}{$M$}
\\
\multicolumn{1}{c}{}
&\multicolumn{1}{c}{}
&\multicolumn{1}{c}{$h^{-1}\,Mpc$}
&\multicolumn{1}{c}{$km\,s^{-1}$}
&\multicolumn{2}{c}{$h^{-1}\,Mpc$}
&\multicolumn{1}{c}{}
&\multicolumn{2}{c}{$h^{-1}\,10^{14}M_{\odot}$}
\\
\multicolumn{1}{c}{(1)}
&\multicolumn{1}{c}{(2)}
&\multicolumn{1}{c}{(3)}
&\multicolumn{1}{c}{(4)}
&\multicolumn{1}{c}{(5)}
&\multicolumn{1}{c}{(6)}
&\multicolumn{1}{c}{(7)}
&\multicolumn{1}{c}{(8)}
&\multicolumn{1}{c}{(9)}
\\
\hline 
A115S\tablenotemark{a,c}              &   13&  0.91& 1074$^{+  208}_{-  121}$&1.40& 0.98&B&12.40$^{+ 5.72}_{- 4.17}$& $\;\;$9.98$^{+ 4.60}_{- 3.36}$\\
A140\tablenotemark{a,c,d}             &    7&  0.13& $\;\;$941$^{+  369}_{-  251}$&1.28& 0.91&-& $\;\;$8.86$^{+ 7.29}_{- 5.22}$& $\;\;$7.11$^{+ 5.85}_{- 4.19}$\\
A222                                  &   26&  0.75&  $\;\;$730$^{+  102}_{-   96}$& 0.93& 0.70&A& $\;\;$4.08$^{+ 1.53}_{- 1.48}$& $\;\;$2.23$^{+  .84}_{-  .81}$\\
A223\tablenotemark{a,c}               &   14&  0.81&  $\;\;$868$^{+  186}_{-  124}$&1.11& 0.81&A& $\;\;$6.67$^{+ 3.31}_{- 2.53}$& $\;\;$3.73$^{+ 1.85}_{- 1.42}$\\
A370                                  &   35&  0.81&  $\;\;$859$^{+  118}_{-  112}$& 0.91& 0.68&C& $\;\;$5.53$^{+ 2.06}_{- 2.00}$& $\;\;$4.75$^{+ 1.76}_{- 1.72}$\\
A520\tablenotemark{a,c}               &   18&  0.60& 1005$^{+  229}_{-  132}$&1.30& 0.92&C&10.23$^{+ 5.32}_{- 3.71}$& $\;\;$8.85$^{+ 4.60}_{- 3.21}$\\
A521                                  &   35&  0.64& 1123$^{+  146}_{-  102}$&1.37& 0.97&B&13.35$^{+ 4.82}_{- 4.13}$&10.74$^{+ 3.87}_{- 3.32}$\\
A665\tablenotemark{c}                 &   25& 2.41&  $\;\;$821$^{+  233}_{-  130}$&1.09& 0.80&-& $\;\;$5.88$^{+ 3.65}_{- 2.37}$& $\;\;$4.70$^{+ 2.92}_{- 1.90}$\\
A851                                  &   55& 1.01& 1067$^{+   89}_{-   96}$&1.09& 0.80&C& $\;\;$9.94$^{+ 2.99}_{- 3.06}$& $\;\;$8.57$^{+ 2.58}_{- 2.64}$\\
A1300                                 &   53&  0.86& 1034$^{+   89}_{-  104}$&1.18& 0.85&B& $\;\;$9.95$^{+ 3.02}_{- 3.19}$& $\;\;$7.97$^{+ 2.42}_{- 2.56}$\\
A2218                                 &   43&  0.32& 1222$^{+  147}_{-  109}$&1.63&1.12&-&18.27$^{+ 6.34}_{- 5.61}$&14.77$^{+ 5.12}_{- 4.54}$\\
A2390                                 &  200& 3.07& 1294$^{+   76}_{-   67}$&1.62&1.11&A&20.35$^{+ 5.62}_{- 5.51}$&11.79$^{+ 3.26}_{- 3.19}$\\
A3639\tablenotemark{a,c,d}            &    7&  0.23&  $\;\;$659$^{+  367}_{-  216}$& 0.91& 0.69&C& $\;\;$3.27$^{+ 3.73}_{- 2.29}$& $\;\;$2.81$^{+ 3.20}_{- 1.97}$\\
A3888                                 &   50& 1.44& 1102$^{+  137}_{-  107}$&1.52&1.05&A&14.00$^{+ 4.94}_{- 4.43}$& $\;\;$8.07$^{+ 2.85}_{- 2.56}$\\
A3889a\tablenotemark{a,b}             &    7&  0.40&    $\;\;\;\;\;\;$0                    & -  &  - &-&  -                      & -                       \\
A3889b\tablenotemark{a,b}             &    9&  0.26&  $\;\;$138$^{+   25}_{-  132}$& 0.17& 0.17&-&  $\;\;$0.04$^{+  .02}_{-  .07}$&  $\;\;$0.02$^{+  .01}_{-  .05}$\\
AS506\tablenotemark{a,c,e}              &   21&  0.38& 1356$^{+  204}_{-  150}$&1.52&1.05&B&21.23$^{+ 8.30}_{- 7.09}$&17.13$^{+ 6.70}_{- 5.72}$\\
AS910                                 &   53&  0.80& 1010$^{+   94}_{-   73}$&1.15& 0.83&B& $\;\;$9.32$^{+ 2.90}_{- 2.69}$& $\;\;$7.45$^{+ 2.32}_{- 2.15}$\\
AS1077                                &   63&  0.67& 1388$^{+  128}_{-   71}$&1.57&1.08&B&22.79$^{+ 7.08}_{- 6.16}$&18.41$^{+ 5.72}_{- 4.97}$\\
CL0017-20\tablenotemark{a,c,e}        &   20&  0.23& 1197$^{+  222}_{-  125}$&1.42& 0.99&-&15.62$^{+ 6.99}_{- 5.09}$&12.58$^{+ 5.63}_{- 4.10}$\\
CLJ0023+0423a\tablenotemark{d}        &    5& 0.48&  $\;\;$283$^{+   53}_{-   17}$& 0.19& 0.19&-&  $\;\;$0.17$^{+  .07}_{-  .05}$& $\;\;$0.11$^{+  .05}_{-  .03}$\\
CLJ0023+0423b\tablenotemark{b,d}      &    5&  0.12&  $\;\;$253$^{+  135}_{-   17}$& 0.17& 0.17&-&  $\;\;$0.12$^{+  .13}_{-  .03}$&  $\;\;$0.08$^{+  .09}_{-  .02}$\\
CL0024+16                             &   73& 1.03&  $\;\;$911$^{+   81}_{-  107}$& 0.94& 0.71&A& $\;\;$6.42$^{+ 1.97}_{- 2.20}$& $\;\;$3.53$^{+ 1.08}_{- 1.21}$\\
CL0053-37\tablenotemark{a,c}          &   20&  0.23& 1136$^{+  259}_{-  167}$&1.54&1.06&-&15.03$^{+ 7.81}_{- 5.80}$&12.13$^{+ 6.31}_{- 4.68}$\\
CL0054-27\tablenotemark{a,c,e}        &    7&  0.46&  $\;\;$742$^{+  599}_{-  147}$& 0.65& 0.52&A& $\;\;$3.12$^{+ 5.10}_{- 1.46}$& $\;\;$1.62$^{+ 2.64}_{-  .76}$\\
CL0303+17                             &   29& 1.04&  $\;\;$876$^{+  144}_{-  140}$& 0.88& 0.67&B& $\;\;$5.62$^{+ 2.32}_{- 2.28}$& $\;\;$4.45$^{+ 1.84}_{- 1.81}$\\
CL0412-65\tablenotemark{a,c,d}        &    6& 1.04&  $\;\;$681$^{+  256}_{-  185}$& 0.62& 0.50&-& $\;\;$2.55$^{+ 2.02}_{- 1.53}$& $\;\;$1.99$^{+ 1.58}_{- 1.19}$\\
CL0949+44a                            &   14&  0.40&  $\;\;$458$^{+  134}_{-  131}$& 0.48& 0.41&A&  $\;\;$0.93$^{+  .59}_{-  .58}$&  $\;\;$0.45$^{+  .29}_{-  .28}$\\
CL0949+44b                            &    8&  0.35&  $\;\;$434$^{+  111}_{-   93}$& 0.47& 0.40&-&  $\;\;$0.82$^{+  .47}_{-  .41}$&  $\;\;$0.63$^{+  .36}_{-  .31}$\\
CL1447+26\tablenotemark{a,c,d}        &    5&  0.67&  $\;\;$838$^{+  163}_{-    1}$& 0.88& 0.67&-& $\;\;$5.15$^{+ 2.38}_{- 1.29}$& $\;\;$4.08$^{+ 1.89}_{- 1.02}$\\
CL1601+42                             &   46&  0.62&  $\;\;$646$^{+   84}_{-   87}$& 0.57& 0.47&C& $\;\;$2.14$^{+  .77}_{-  .79}$& $\;\;$1.81$^{+  .65}_{-  .67}$\\
CLJ1604+4304\tablenotemark{c}         &    8&  0.36&  $\;\;$858$^{+  277}_{-   83}$& 0.56& 0.46&-& $\;\;$3.68$^{+ 2.55}_{- 1.16}$& $\;\;$2.85$^{+ 1.97}_{-  .90}$\\
F1637.23TL\tablenotemark{a,c,d}  &    6&  0.35&  $\;\;$538$^{+  106}_{-  367}$& 0.51& 0.42&-& $\;\;$1.34$^{+  .63}_{- 1.87}$& $\;\;$1.03$^{+  .48}_{- 1.43}$\\
F1652.20CR\tablenotemark{a,b,c,d}&    6&  0.25&  $\;\;$510$^{+  511}_{-  511}$& 0.52& 0.43&-& $\;\;$1.23$^{+ 2.48}_{- 2.48}$&  $\;\;$0.94$^{+ 1.91}_{- 1.91}$\\
J2175.15TR\tablenotemark{a,b,c,d}&    8&  0.36&  $\;\;$246$^{+   79}_{-  239}$& 0.25& 0.24&-&  $\;\;$0.16$^{+  .11}_{-  .31}$&  $\;\;$0.11$^{+  .08}_{-  .22}$\\
J2175.23C\tablenotemark{a,c,d}   &    5&  0.29&  $\;\;$443$^{+  177}_{-  430}$& 0.45& 0.38&-&  $\;\;$0.83$^{+  .69}_{- 1.62}$&  $\;\;$0.63$^{+  .53}_{- 1.23}$\\
MS0015.9+1609                    &   42& 1.14&  $\;\;$984$^{+  130}_{-   95}$& 0.87& 0.66&A& $\;\;$7.00$^{+ 2.55}_{- 2.21}$& $\;\;$3.80$^{+ 1.38}_{- 1.20}$\\
MS0302.7+1658                    &   30&  0.86&  $\;\;$735$^{+  109}_{-   80}$& 0.73& 0.58&A& $\;\;$3.40$^{+ 1.32}_{- 1.13}$& $\;\;$1.80$^{+  .70}_{-  .60}$\\
MS0302.5+1717                    &   24&  0.41&  $\;\;$664$^{+   67}_{-   77}$& 0.66& 0.53&C& $\;\;$2.56$^{+  .82}_{-  .87}$& $\;\;$2.17$^{+  .70}_{-  .74}$\\
MS0440.5+0204                    &   32&  0.23&  $\;\;$838$^{+  131}_{-  139}$&10.09& 0.80&B& $\;\;$6.13$^{+ 2.45}_{- 2.55}$& $\;\;$4.90$^{+ 1.96}_{- 2.03}$\\
MS0451.6-0305                    &   40&  0.99& 1317$^{+  122}_{-  103}$&1.17& 0.85&A&16.10$^{+ 5.01}_{- 4.75}$& $\;\;$9.06$^{+ 2.82}_{- 2.67}$\\
MS1008.1-1224                    &   65&  0.77& 1033$^{+  115}_{-  105}$&1.18& 0.85&C& $\;\;$9.93$^{+ 3.33}_{- 3.20}$& $\;\;$8.58$^{+ 2.87}_{- 2.76}$\\
MS1054.4-0321                    &   32&  0.72& 1178$^{+  139}_{-  113}$& 0.81& 0.62&A& $\;\;$9.46$^{+ 3.25}_{- 2.98}$& $\;\;$5.08$^{+ 1.75}_{- 1.60}$\\
MS1224.7+2007                    &   23&  0.86&  $\;\;$837$^{+  100}_{-   83}$& 0.93& 0.70&A& $\;\;$5.38$^{+ 1.86}_{- 1.72}$& $\;\;$2.95$^{+ 1.02}_{-  .94}$\\
MS1358.4+6245                    &  133& 1.16&  $\;\;$985$^{+   58}_{-   62}$&10.09& 0.80&B& $\;\;$8.51$^{+ 2.35}_{- 2.38}$& $\;\;$6.80$^{+ 1.88}_{- 1.90}$\\
MS1512.4+3647                    &   35& 2.20&  $\;\;$776$^{+  172}_{-  103}$& 0.82& 0.63&C& $\;\;$4.16$^{+ 2.12}_{- 1.52}$& $\;\;$3.56$^{+ 1.81}_{- 1.30}$\\
MS1621.5+2640                    &   88& 2.51&  $\;\;$735$^{+   53}_{-   53}$& 0.73& 0.57&A& $\;\;$3.40$^{+  .98}_{-  .98}$& $\;\;$1.80$^{+  .52}_{-  .52}$\\
RXJ1716+67\tablenotemark{c}      &   19& 1.03& 1445$^{+  288}_{-  218}$&1.01& 0.75&A&17.17$^{+ 8.08}_{- 6.73}$& $\;\;$9.51$^{+ 4.47}_{- 3.73}$\\
1E0657-56                        &   12&  0.59&  $\;\;$926$^{+  178}_{-  104}$&10.07& 0.78&B& $\;\;$7.36$^{+ 3.38}_{- 2.47}$& $\;\;$5.87$^{+ 2.69}_{- 1.97}$\\
3C206\tablenotemark{a,c,d}       &    7&  0.22&  $\;\;$585$^{+  574}_{-  155}$& 0.76& 0.59&-& $\;\;$2.21$^{+ 4.38}_{- 1.30}$& $\;\;$1.74$^{+ 3.45}_{- 1.02}$\\
3C295\tablenotemark{c,e}         &   15&  0.36& 1642$^{+  224}_{-  187}$&1.58&1.09&B&32.22$^{+11.92}_{-10.90}$&26.03$^{+ 9.63}_{- 8.80}$\\
\hline
\end{tabular}

%% file: comm_tab3.tex
{\footnotesize\parindent=3mm
$^a$~
Clusters having in their field less than 30 galaxies with available redshift
(cf. Table~1).\\ 
$^b$~
Clusters with a peak in the
velocity distribution less significant that $99\%$.\\
$^c$~
Clusters with an
error on $\sigma_v$ of $\gtrsim 150$ \kss.\\
$^d$~
Clusters with a VDP which is
poorly defined.\\
$^e$~
Clusters with a VDP which is
without a flat behavior in the external cluster
regions: the strong decreasing suggests that the estimates of 
$\sigma_v$,\\
 $M_V$,
and $M$ are better interpreted as upper limits (see text).
}

%% file: tab4.tex
%
%

\begin{tabular}{lcclcl}
\hline\hline
\multicolumn{1}{c}{Cluster Name}
&\multicolumn{1}{c}{$L_{X,bol}$}
&\multicolumn{1}{c}{References}
&\multicolumn{1}{c}{$T_X$}
&\multicolumn{1}{c}{References}
&\multicolumn{1}{c}{$\beta_{spec}$}
\\
\multicolumn{1}{c}{}
&\multicolumn{1}{c}{$h^{-2}\,10^{44}\,erg\,s^{-1}$}
&\multicolumn{1}{c}{}
&\multicolumn{1}{c}{$keV$}
&\multicolumn{1}{c}{}
&\multicolumn{1}{c}{}
\\
\multicolumn{1}{c}{(1)}
&\multicolumn{1}{c}{(2)}
&\multicolumn{1}{c}{(3)}
&\multicolumn{1}{c}{(4)}
&\multicolumn{1}{c}{(5)}
&\multicolumn{1}{c}{(6)}
\\
\hline 
A115S               &   2.26375&1&    -   &-&    -   \\
A222                &   1.9125&2&    -   &-&    -   \\
A223\tablenotemark{a}                &    0.7&3& -   &-&    -   \\
A370\tablenotemark{b}                &   5.1925&2&    $\;\;$7.13$^{+    1.05}_{-     .83}$
&2&     0.63$^{+     .18}_{-     .17}$\\
A520\tablenotemark{b}                &   9.3375&2&    $\;\;$8.59$^{+     .93}_{-     .93}$&2&     0.71$^{+     .33}_{-     .19}$\\
A521\tablenotemark{a}                &   2.72&4& - &-&    -   \\
A665\tablenotemark{b}&       10.43&2&    $\;\;$8.26$^{+     .95}_{-     .81}$&2&     0.49$^{+     .28}_{-     .16}$\\
A851\tablenotemark{b}                &   4.02&2&    $\;\;$$\;\;$6.7$^{+    2.7}_{-    1.7}$&2&    1.03$^{+     .31}_{-     .25}$\\
A1300\tablenotemark{b}               &  11.9075&2&   $\;\;$11.4$^{+    1.3}_{-    1.0}$&11&     0.57$^{+     .11}_{-     .12}$\\
A2218\tablenotemark{b}               &   5.49&2&    $\;\;$$\;\;$7.05$^{+     .36}_{-     .35}$&2&    1.28$^{+     .31}_{-     .23}$\\
A2390\tablenotemark{b}               &  15.8725&2&    $\;\;$11.1$^{+    1.5}_{-    1.6}$&2&     0.91$^{+     .13}_{-     .13}$\\
A3888               &   7.85&2&    -   &-&    -   \\
AS506\tablenotemark{b}               &   4.3775&2&    $\;\;$$\;\;$7.2$^{+    3.7}_{-    1.8}$&2&    1.55$^{+    .68}_{-     .42}$\\
AS1077\tablenotemark{b}              &   9.525&2&    $\;\;$9.76$^{+    1.04}_{-     .85}$&2&    1.20$^{+     .23}_{-     .14}$\\
CL0024+16\tablenotemark{b}           &   1.5225&5&    $\;\;$$\;\;$5.7$^{+    4.9}_{-    2.1}$&5&     0.88$^{+     .50}_{-     .29}$\\
CL0054-27\tablenotemark{a}           &    0.4325&6&  - &-&    -   \\
CL0303+17           &    0.9&7&    -   &-&    -   \\
CL0412-65\tablenotemark{a}           &    0.1325&6&  -  &-&    -   \\
CL1447+26           &   2.6725&2&    -   &-&    -   \\
CLJ1604+4304\tablenotemark{a}        &    0.535&8& -   &-&    -   \\
1E0657-56\tablenotemark{b}           &  30.&9&   $\;\;$11.7$^{+    2.2}_{-    1.4}$&12&     0.44$^{+     .18}_{-     .11}$\\
F1637.23TL\tablenotemark{a}          &    0.2725&10& -  &-&    -   \\
F1652.20CR\tablenotemark{a}          &    0.2225&10& -  &-&    -   \\
J2175.15TR\tablenotemark{a}          &    0.3625&10& -  &-&    -   \\
J2175.23C\tablenotemark{a}           &    0.0875&10& -  &-&    -   \\
MS0015.9+1609\tablenotemark{b}       &   7.0325&2&    $\;\;$$\;\;$8.0$^{+    1.0}_{-    1.0}$&2&     0.73$^{+     .20}_{-     .15}$\\
MS0302.7+1658       &   2.27&2&    $\;\;$$\;\;$4.6$^{+     .8}_{-     .8}$&2&     0.71$^{+     .24}_{-     .20}$\\
MS0302.5+1717       &   1.0575&2&    -   &-&    -   \\
MS0440.5+0204\tablenotemark{b}       &   1.857&2&    $\;\;$5.30$^{+    1.27}_{-     .85}$&2&     0.80$^{+     .28}_{-     .28}$\\
MS0451.6-0305       &   3.9825&2&   10.17$^{+    1.55}_{-    1.26}$&2&    1.03$^{+     .22}_{-     .18}$\\
MS1008.1-1224\tablenotemark{b}       &   2.2825&2&    $\;\;$7.29$^{+    2.45}_{-    1.52}$&2&     0.89$^{+     .27}_{-     .21}$\\
MS1054.4-0321\tablenotemark{b}       &   4.9775&2&   $\;\;$12.3$^{+    3.1}_{-    2.2}$&2&     0.68$^{+     .19}_{-     .15}$\\
MS1224.7+2007       &   2.015&2&    $\;\;$$\;\;$4.3$^{+     .7}_{-     .6}$&2&     0.99$^{+     .29}_{-     .24}$\\
MS1358.4+6245\tablenotemark{b}       &   5.4525&2&    $\;\;$$\;\;$7.5$^{+    7.1}_{-    1.5}$&2&     0.78$^{+     .47}_{-     .14}$\\
MS1512.4+3647\tablenotemark{b}       &   1.905&2&    $\;\;$3.57$^{+    1.33}_{-     .64}$&2&    1.02$^{+     .51}_{-     .29}$\\
MS1621.5+2640       &   2.055&2&    -   &-&    -   \\
RXJ1716+67          &   4.35&2&    $\;\;$5.66$^{+    1.37}_{-     .58}$&2&    2.24$^{+    1.04}_{-     .71}$\\
3C295\tablenotemark{b}               &   6.5&2&    $\;\;$7.13$^{+    2.06}_{-    1.35}$&2&    2.29$^{+    .75}_{-    .59}$\\
\hline
\end{tabular}

%% file: comm_tab4.tex
{\footnotesize\parindent=3mm
$^a$~
Bolometric luminosity here computed from  the original band
luminosity.\\
$^a$~
Errors on $T_X$ are  at the $90\%$ c.l.; they are rescaled
by a factor 1.6 to compute $68\%$ c.l. errors on $\beta_{spec}$.\\
(1) White et al. 1997; (2) Wu et al. 1999 (see this compilation
for original data sources); (3) Lea \& Henry 1988; (4) Ulmer et al. 1985;
(5) Soucail \\
et al. 2000; (6) Smail et al. 1997; (7) Kaiser et al. 1998;
(8) Castander et al. 1994; (9) Tucker et al. 1998; (10) Bower et al. 1997;
(11) Pierre et \\ al. (1999); (12) Yaqoob 1999. 
}

%% file: tab5.tex
%
%

\begin{tabular}{lcclll}
\hline \hline
\multicolumn{1}{c}{Name}
&\multicolumn{1}{c}{References}
&\multicolumn{1}{c}{$R_L$}
&\multicolumn{1}{c}{$M_L$}
&\multicolumn{1}{c}{$M_{opt,L}$}
&\multicolumn{1}{c}{$M_{opt,L}/M_L$}
\\
\multicolumn{1}{c}{}
&\multicolumn{1}{c}{}
&\multicolumn{1}{c}{$h^{-1}\,Mpc$}
&\multicolumn{1}{c}{$h^{-1}\,10^{14}M_{\odot}$}
&\multicolumn{1}{c}{$h^{-1}\,10^{14}M_{\odot}$}
&\multicolumn{1}{c}{}
\\
\multicolumn{1}{c}{(1)}
&\multicolumn{1}{c}{(2)}
&\multicolumn{1}{c}{(3)}
&\multicolumn{1}{c}{(4)}
&\multicolumn{1}{c}{(5)}
&\multicolumn{1}{c}{(6)}
\\
\hline
A851                &1&    0.375&   $\;\;$$\;\;$$\;\;$3.0$^{+     .5}_{-     .5}$&    3.792$^{+    1.140}_{-    1.168}$&     1.26$^{+      .43}_{-      .44}$\\
A2218               &2&    0.4&    $\;\;$$\;\;$$\;\;$3.9$^{+     .7}_{-     .7}$&    5.206$^{+    1.806}_{-    1.599}$&     1.33$^{+      .52}_{-      .47}$\\
A2390               &3&    0.59&   $\;\;$$\;\;$10.0$^{+    3.5}_{-    3.6}$&    5.697$^{+    1.574}_{-    1.542}$&      0.57$^{+      .25}_{-      .25}$\\
AS1077              &4&    0.25&    $\;\;$$\;\;$$\;\;$2.0$^{+     .2}_{-     .2}$&    4.424$^{+    1.374}_{-    1.195}$&     2.21$^{+      .72}_{-      .64}$\\
CL0024+16           &5&    0.2&    $\;\;$$\;\;$1.38$^{+     .37}_{-     .37}$&     0.979$^{+     .300}_{-     .336}$&      0.71$^{+      .29}_{-      .31}$\\
CL0054-27           &5&    0.2&    $\;\;$$\;\;$1.71$^{+     .64}_{-     .64}$&     0.593$^{+     .969}_{-     .278}$&      0.35$^{+      .58}_{-      .21}$\\
CL0303+17           &6&    0.592&     $\;\;$0.981$^{+     .312}_{-     .312}$&    3.307$^{+    1.366}_{-    1.342}$&     3.37$^{+     1.76}_{-     1.74}$\\
CL0412-65           &5&    0.2&     $\;\;$$\;\;$0.25$^{+     .41}_{-     .41}$&     0.749$^{+     .593}_{-     .448}$&     3.00$^{+     5.46}_{-     5.23}$\\
CL1601+42           &5&    0.2&     $\;\;$$\;\;$0.77$^{+     .66}_{-     .66}$&     0.729$^{+     .263}_{-     .268}$&      0.95$^{+      .88}_{-      .88}$\\
MS0015.9+1609       &5&    0.2&    $\;\;$$\;\;$1.87$^{+     .64}_{-     .64}$&    1.120$^{+     .408}_{-     .354}$&      0.60$^{+      .30}_{-      .28}$\\
MS0302.7+1658       &6&    0.596&     $\;\;$0.624$^{+     .315}_{-     .315}$&    1.539$^{+     .597}_{-     .510}$&     2.47$^{+     1.57}_{-     1.49}$\\
MS0302.5+1717       &6&    0.595&    $\;\;$2.069$^{+     .387}_{-     .387}$&    1.998$^{+     .642}_{-     .681}$&      0.97$^{+      .36}_{-      .38}$\\
MS1008.1-1224       &7&    0.34&    $\;\;$$\;\;$2.18$^{+     .47}_{-     .47}$&    3.308$^{+    1.108}_{-    1.066}$&     1.52$^{+      .60}_{-      .59}$\\
MS1054.4-0321       &8&    0.8&    $\;\;$$\;\;$$\;\;$$\;\;$8.$^{+   21.}_{-    5.}$&    5.049$^{+    1.736}_{-    1.591}$&      0.63$^{+     1.67}_{-      .83}$\\
MS1224.7+2007       &9&    0.65&    $\;\;$$\;\;$$\;\;$4.7$^{+    2.0}_{-    1.5}$&    2.258$^{+     .781}_{-     .721}$&      0.48$^{+      .26}_{-      .23}$\\
MS1358.4+6245       &10&    0.5&    $\;\;$$\;\;$$\;\;$2.2$^{+     .3}_{-     .3}$&    3.791$^{+    1.048}_{-    1.061}$&     1.72$^{+      .53}_{-      .54}$\\
RXJ1716+67          &11&    0.5&    $\;\;$$\;\;$$\;\;$2.6$^{+     .9}_{-     .9}$&    5.607$^{+    2.638}_{-    2.197}$&     2.16$^{+     1.26}_{-     1.13}$\\
3C295               &5&    0.2&    $\;\;$$\;\;$2.35$^{+     .38}_{-     .38}$&    5.014$^{+    1.855}_{-    1.696}$&     2.13$^{+      .86}_{-      .80}$\\
\hline
\end{tabular}

%% file: comm_tab5.tex
{\footnotesize\parindent=3mm
(1) Seitz et al. 1996; (2) Squires et al.  1996a;
(3) Squires et al. 1996b; (4) Natarajan et al. 1998;
(5) Smail et al. 1997; (6) Kaiser et al. 1998; \\
(7) Athreya et al.
1999, see also Lombardi et al. 2000; 
(8) Luppino \& Kaiser 1997; (9) Fischer 1999; (10) Hoekstra et al.
1998;
(11) Clowe et al. \\ 1998.
}

%% file: tab6.tex
%
%

\begin{tabular}{lrrlrrcll}
\hline \hline
\multicolumn{1}{c}{Name}
&\multicolumn{1}{c}{$N_m$}
&\multicolumn{1}{c}{$R_{max}$}
&\multicolumn{1}{c}{$\sigma_v$}
&\multicolumn{1}{c}{$R_{vir}$}
&\multicolumn{1}{c}{$R_{PV}$}
&\multicolumn{1}{c}{$T$}
&\multicolumn{1}{c}{$M_V$}
&\multicolumn{1}{c}{$M$}
\\
\multicolumn{1}{c}{}
&\multicolumn{1}{c}{}
&\multicolumn{1}{c}{$h^{-1}\,Mpc$}
&\multicolumn{1}{c}{$km\,s^{-1}$}
&\multicolumn{2}{c}{$h^{-1}\,Mpc$}
&\multicolumn{1}{c}{}
&\multicolumn{2}{c}{$h^{-1}\,10^{14}M_{\odot}$}
\\
\multicolumn{1}{c}{(1)}
&\multicolumn{1}{c}{(2)}
&\multicolumn{1}{c}{(3)}
&\multicolumn{1}{c}{(4)}
&\multicolumn{1}{c}{(5)}
&\multicolumn{1}{c}{(6)}
&\multicolumn{1}{c}{(7)}
&\multicolumn{1}{c}{(8)}
&\multicolumn{1}{c}{(9)}
\\
\hline 
A1689a              &   38& 2.26&  $\;\;$765$^{+   80}_{-   60}$&1.01& 0.75&C& $\;\;$4.81$^{+ 1.57}_{- 1.42}$& $\;\;$4.14$^{+ 1.35}_{- 1.22}$\\
A1689b              &   15& 1.67&  $\;\;$636$^{+  167}_{-  127}$& 0.85& 0.65&B& $\;\;$2.87$^{+ 1.67}_{- 1.35}$& $\;\;$2.27$^{+ 1.32}_{- 1.07}$\\
A1689ab             &   49& 2.26& 1172$^{+  123}_{-   99}$&1.55&1.07&B&16.12$^{+ 5.26}_{- 4.86}$&13.02$^{+ 4.25}_{- 3.93}$\\
A2744a              &   34&  0.53& 1121$^{+  176}_{-   88}$&1.28& 0.91&C&12.59$^{+ 5.05}_{- 3.72}$&10.90$^{+ 4.37}_{- 3.22}$\\
A2744b              &   25&  0.64&  $\;\;$682$^{+   97}_{-   75}$& 0.77& 0.60&C& $\;\;$3.04$^{+ 1.15}_{- 1.01}$& $\;\;$2.60$^{+  .98}_{-  .87}$\\
A2744ab             &   55&  0.59& 1777$^{+  151}_{-  125}$&2.02&1.34&B&46.36$^{+14.01}_{-13.30}$&37.66$^{+11.38}_{-10.80}$\\
A3854a              &   18&  0.70&  $\;\;$455$^{+   43}_{-  102}$& 0.63& 0.50&-& $\;\;$1.14$^{+  .36}_{-  .59}$&  $\;\;$0.89$^{+  .28}_{-  .46}$\\
A3854b              &    9&  0.75&  $\;\;$520$^{+  163}_{-  254}$& 0.72& 0.57&-& $\;\;$1.68$^{+ 1.13}_{- 1.69}$& $\;\;$1.32$^{+  .89}_{- 1.33}$\\
A3854ab             &   30&  0.81& 1211$^{+  210}_{-  138}$&1.67&1.14&A&18.31$^{+ 7.83}_{- 6.19}$&10.63$^{+ 4.55}_{- 3.60}$\\
\hline
\end{tabular}